\documentclass[11pt,a4paper]{article}
\usepackage[utf8]{inputenc}
\usepackage{amsmath}
\usepackage{amsfonts}
\usepackage{amssymb}
\usepackage{graphicx}
\usepackage{authblk}
\usepackage{csvsimple} 
\usepackage{booktabs}
\usepackage{url} 
\usepackage{longtable} 
\usepackage{xcolor} 

\title {\textbf{Sparse Correspondence Analysis\\
 for Contingency Tables}}

\author[1]{Ruiping Liu}
\author[2]{Ndeye Niang}
\author[2]{Gilbert Saporta\thanks{Corresponding author: \texttt{gilbert.saporta@cnam.fr}}}
\author[3]{Huiwen Wang}
 
\affil[1]{Beijing Information Science \& Technology University and Beihang University, Beijing, China}
\affil[2]{CEDRIC Lab, CNAM, Paris, France}
\affil[3]{School of Economics and Management and Beijing Advanced Innovation Center for Big Data \& Brain Computing, Beihang University, Beijing, China}

\begin{document}

\hyphenation{corres-pondances}

\maketitle

\noindent \textbf{Abstract} Since the introduction of the lasso in regression, various sparse methods have been developed in an unsupervised context like sparse principal component analysis (s-PCA), sparse canonical correlation analysis \linebreak (s-CCA) and sparse singular value decomposition (s-SVD).  These sparse methods combine feature selection and dimension reduction. One advantage of s-PCA is to simplify the interpretation of the (pseudo) principal components since each one is expressed as a linear combination of a small number of variables. The disadvantages lie on the one hand in the difficulty of choosing the number of non-zero coefficients in the absence of a well established criterion and on the other hand in the loss of orthogonality for the components and/or the loadings.  
 
In this paper we propose sparse variants of correspondence analysis (CA) for large contingency tables like documents-terms matrices used in text mining, together with pPMD, a deflation technique derived from projected deflation in s-PCA. We use the fact that CA is a double weighted PCA (for rows and columns) or a weighted SVD, as well as a canonical correlation analysis of indicator variables.  Applying s-CCA or s-SVD allows to sparsify both rows and columns weights. The user may tune the level of sparsity of rows and columns and optimize it according to some criterium, and even decide that no sparsity is needed for rows (or columns) by relaxing one sparsity constraint.   The latter is equivalent to apply s-PCA to matrices of row (or column) profiles.

\noindent \textbf{Keywords} sparse methods, correspondence analysis, textual data, sparse correspondence analysis

\section{Introduction and motivation}

When the number of columns (or rows) in a contingency table is very high, as in textual data (rows = documents, columns = terms), the interpretation of CA is difficult. Hence the idea of a sparse version of CA need to be considered, where the factor scores are sparse, i.e. contain many zero coefficients. 

CA of a contingency table is a double weighted PCA of rows and columns profiles with the chi-square distance, or a weighted SVD. CA is also a canonical correlation analysis (CCA) between the two groups of indicator variables of the row and column categories and lastly CA is a particular Multiple Correspondence Analysis (MCA) with $p$ = 2 variables. 
Therefore applying sparse versions of PCA, SVD, and CCA leads to sparse versions of CA but sparse CA cannot be a particular case of sparse Multiple Correspondence Analysis as it will be shown later.

As CA treats symmetrically rows and columns, it is natural to develop what we call a double sparse CA where both rows and columns weights include many zeros.  However, when one is not interested in sparsifying both rows and columns, a column sparse only CA comes down to a sparse PCA of the row profile matrix with weighted rows and chi-square metric: only columns loadings are sparse. This could be the case in some documents-terms analysis where one looks for components which are explained each by a small number of words, while one keeps all the documents.

Since there are several versions of sparse PCA, CCA, and SVD with different properties, a choice should be done. Like in other sparse methods some properties may be lost such as orthogonality and barycentric relations in CA.

The rest of the paper is organized as follows. Section 2 gives a reminder on CA. The various sparse versions of PCA, CCA, MCA, SVD as well as  general related issues are presented in section 3 and their main properties are compared. The proposed sparse CA method is detailed in section 4 together with a toy example. Applications of sparse CA to textual data are presented in section 5. Conclusion and perspectives for future work are given in section 6.

\section{Reminder on CA }

According to Abdi \&  Béra (2017), “\textit{CA is traditionally used to analyze contingency tables. CA decomposes the chi-square statistics associated to the data table into two sets of orthogonal components that describe, respectively, the pattern of associations between the elements of the rows and between the elements of the columns of the data table. When the data table is a set of observations described by a set of nominal variables, CA becomes multiple correspondence analysis (MCA)}.”  

A detailed presentation can be found in Greenacre (2010). Lebart \& Saporta (2014) propose an historical sketch. One can also refer to the comprehensive book  of  Beh \& Lombardo (2014).

Notations are as follows:  bold uppercase letters like ${\bf{X}}$ denote matrices, bold lowercase letters like $ \textbf{x} $ denote vectors. Italic lower case letters denote scalars or elements. In PCA $ n$ is the number of observations, $ p $ the number of centered variables, $\bf{v} $  a norm 1  eigenvector of  principal coefficients or a right singular vector of ${\bf{X}}$, $ \textbf{ z}$ a principal component such as $ \textbf{z}=\textbf{Xv} $,  $ \textbf{u} $ a left singular vector of $ \textbf{X} $ , ${\bf{\Sigma }}$  the variance covariance matrix associated to ${\bf{X}}$ :  ${\bf{\Sigma }} = {1 \over n}{\bf{X'X}}$ where superscript $^{'}$ denotes matrix transposition.
In correspondence analysis principal components (rows and columns coordinates) will be denoted $ \textbf{a} $  and $ \textbf{b} $ .
The operation transforming a vector into a diagonal matrix is denoted  $diag$  .  The symbol $\propto$ means proportional to.

\subsection{CA standard equations}
Given two categorical variables with $I$ and $J$ categories, the contingency table $\textbf{N}$ is first transformed into the frequency table $\textbf{P}=\textbf{N}/n$, where $n$ is the total of the $I{\times}J$ elements of $\textbf{N}$.  Let $\textbf{r}$ and $\textbf{c}$ be the vectors containing respectively the rows and columns totals of $\textbf{P}$ (sum=1), in other terms the marginal probabilities $p_{i}$ and $p_{j}$ .  

${{\bf{D}}_r}$ and ${{\bf{D}}_c}$ are the associated diagonal matrices ${{\bf{D}}_r}=diag(\textbf{r})$ and ${{\bf{D}}_c}=diag(\textbf{c})$. ${\bf{D}}_r^{ - 1}{\bf{P}}$ is the matrix of row profiles (row conditional distributions) and ${\bf{D}}_c^{ - 1}{\bf{P}}'$  is the transposed matrix of the column profiles.

The row coordinates (principal component or row scores) are the eigenvectors $ \textbf{a} $  of  ${\bf{D}}_r^{ - 1}{\bf{PD}}_c^{ - 1}{\bf{P}}'$ with scaling  ${\bf{a}}'{{\bf{D}}_r}{\bf{a}} = \lambda $, while the column coordinates are the eigenvectors $ \textbf{b} $ of  ${\bf{D}}_c^{ - 1}{\bf{P}}'{\bf{D}}_r^{ - 1}{\bf{P}}$ with the same eigenvalue ${\bf{b}}'{{\bf{D}}_c}{\bf{b}} = \lambda $
\begin{equation}
{\bf{D}}_r^{ - 1}{\bf{PD}}_c^{ - 1}{\bf{P}}'{\bf{a}} = \lambda {\bf{a}}
\end{equation}
\begin{equation}
{\bf{D}}_c^{ - 1}{\bf{P}}'{\bf{D}}_r^{ - 1}{\bf{Pb}} = \lambda {\bf{b}}
\end{equation}
Equations (1) and (2) have a trivial extraneous solution $ \lambda=1 $ which has to be discarded. The number of components is $min(I-1,J-1)$.

Note that equations (1) and (2) are not solved independently, which could lead to undesired reversions of principal axis. The so-called transition formulas link $ \textbf{a} $  and $ \textbf{b}$ for a given eigenvalue  $\lambda $:

\begin{equation}
{\bf{a}} = {1 \over {\sqrt \lambda  }}{\bf{D}}_r^{ - 1}{\bf{Pb}}  \quad and  \quad {\bf{b}} = {1 \over {\sqrt \lambda  }}{\bf{D}}_c^{ - 1}{\bf{P}}'{\bf{a}} 
\end{equation}

Vectors $ \textbf{a} $  and $ \textbf{b} $ contains rows and columns coordinates on a common direction. This allows simultaneous mappings of the $I+J  $ categories of both nominal variables, which is one of the major attractions of CA. 
We use here the so-called symmetric map (Greenacre, 2010).  Transition formulas (3) are often interpreted as (pseudo) barycentric properties: the coordinate of a row (a column) is proportional to the weighted mean of the coordinates of all columns (rows), with weights equal to the conditional frequencies of the columns (rows) given that row (column)

The sum of all eigenvalues (or total inertia) is equal to Pearson’s  ${\phi ^2}$ :
\begin{equation}
\sum\limits_{k = 1}^{\min (I - 1,J - 1)} {{\lambda _k}}  = \sum\limits_{i = 1}^I {\sum\limits_{j = 1}^J {{{{{\left( {{p_{ij}} - {p_i}{p_j}} \right)}^2}} \over {{p_i}{p_j}}}} }  = {\phi ^2}
\end{equation}
According to Abdi \& Béra, equation (4) shows that  “\textit{the factors of CA perform an orthogonal decomposition of the independence where each factor “explains” a portion of the deviation to independence}”  .

\subsection{CA as two weighted PCAs with chi-square distance}
Rows and columns principal coordinates may be obtained by performing two different weighted PCAs: the first one on the doubly centered matrix of row profiles ${\bf{D}}_r^{ - 1}\left( {{\bf{P}} - {\bf{rc}}'} \right)$  and the second one on the transposed of the doubly centered matrix of columns profiles  ${\bf{D}}_c^{ - 1}\left( {{\bf{P}} - {\bf{rc}}'} \right)'$ (centering eliminates the trivial solution). 

In the first PCA it is necessary to weight the rows  by matrix  ${\bf{D}}_r^{}$ and use the chi-square metric ${\bf{D}}_c^{ - 1}$ for computing distances between row profiles. This weighted PCA  comes down  to  a standard (uncentered and unscaled) PCA of ${\bf{D}}_r^{ - 1/2}\left( {{\bf{P}} - {\bf{rc}}'} \right){\bf{D}}_c^{ - 1/2}$.
   
Similar results are obtained for PCA of column profiles with column weights  ${\bf{D}}_c^{}$ and chi-square metric ${\bf{D}}_r^{ - 1}$.

One important property is that multiplying by  $\alpha  = \sqrt \lambda  $ the vector of the principal coefficients of one PCA gives the vector of the principal coordinates of the other PCA. 
\subsection{CA as a generalized  SVD}
From the standard equations Eq.(1-2), we have ${\bf{D}}_r^{ - {1 \over 2}}{\bf{PD}}_c^{ - 1}{\bf{P}}'{\bf{a}} = \lambda {\bf{D}}_r^{{1 \over 2}}{\bf{a}}$ which implies that ${\bf{v}} = {\bf{D}}_r^{{1 \over 2}}{\bf{a}}$  is an eigenvector of matrix ${\bf{D}}_r^{ - {1 \over 2}}{\bf{PD}}_c^{ - 1}{\bf{P}}'{\bf{D}}_r^{ - {1 \over 2}}$. 
Considering the scaling condition ${\bf{a}}'{{\bf{D}}_r}{\bf{a}} = \lambda $ the row coordinates vector is ${\bf{a}} = \sqrt \lambda  {\bf{D}}_r^{ - {1 \over 2}}{\bf{u}}$ where $\textbf{u}$ is the normalized eigenvector of ${\bf{D}}_r^{ - {1 \over 2}}{\bf{PD}}_c^{ - 1}{\bf{P}}'{\bf{D}}_r^{ - {1 \over 2}}$. So, we can obtain the solution of CA by computing the eigenvalues and eigenvectors of ${\bf{D}}_r^{ - {1 \over 2}}{\bf{PD}}_c^{ - 1}{\bf{P}}'{\bf{D}}_r^{ - {1 \over 2}}$.

Similarly, we can obtain the column coordinates by ${\bf{b}} = \sqrt \lambda  {\bf{D}}_c^{ - {1 \over 2}}{\bf{v}}$ with $\textbf{v}$ being the normalized eigenvector of matrix ${\bf{D}}_c^{ - {1 \over 2}}{\bf{P}}'{\bf{D}}_r^{ - 1}{\bf{PD}}_c^{ - {1 \over 2}}$.
Discarding the first eigenvalue $ \lambda=1 $ is equivalent to consider the doubly centered matrices:
$${\bf{D}}_r^{ - {1 \over 2}}\left( {{\bf{P}} - {\bf{rc}}'} \right){\bf{D}}_c^{ - 1}\left( {{\bf{P}} - {\bf{rc}}'} \right)'{\bf{D}}_r^{ - {1 \over 2}}$$ and $${\bf{D}}_c^{ - {1 \over 2}}\left( {{\bf{P}} - {\bf{rc}}'} \right)'{\bf{D}}_r^{ - 1}\left( {{\bf{P}} - {\bf{rc}}'} \right){\bf{D}}_c^{ - {1 \over 2}}$$ 
Finally CA is equivalent to a doubly weighted SVD of $\left( {{\bf{P}} - {\bf{rc}}'} \right)$\begin{equation}
{\bf{D}}_r^{ - 1/2}\left( {{\bf{P}} - {\bf{rc}}'} \right){\bf{D}}_c^{ - 1/2} = {\bf{U}}{{\bf{D}}_\alpha }{\bf{V}}'
\end{equation}
where ${{\bf{D}}_\alpha }$ is the diagonal matrix of the singular values $\alpha  = \sqrt \lambda  $.

The left-hand matrix has entries equal to  $${{{p_{ij}} - {p_i}{p_j}} \over {\sqrt {{p_i}{p_j}} }}$$  The matrix $\textbf{A}$ of the row coordinates on principal axes is given by
\begin{equation}
{\bf{A}} = {\bf{D}}_r^{ - 1/2}{\bf{U}}{{\bf{D}}_\alpha }
\end{equation}
and the matrix $\textbf{B}$ of the column coordinates on principal axes is given by 
\begin{equation}
{\bf{B}} = {\bf{D}}_c^{ - 1/2}{\bf{V}}{{\bf{D}}_\alpha }
\end{equation}
It may be noted that the SVD of ${\bf{D}}_r^{ - 1/2}{\bf{PD}}_c^{ - 1/2}$  gives the same results as (5) plus the extraneous or trivial solution corresponding to $\lambda  = 1$.

\subsection{CA as a Canonical Correlation Analysis of indicator matrices}
Canonical correlation analysis (CCA)  aims at studying associations between two sets of variables by maximizing the correlation between linear combinations of the variables in each data set.
 
Let ${{\bf{X}}_1}$ and ${{\bf{X}}_2}$ denote the data tables with $n$ rows and respectively $I$ and $J$ columns.
Without loss of generality,  all variables are assumed to have zero mean. 

${\bf{\Sigma }}$ denotes the variance covariance matrix of  ${{\bf{X}}_1}$ and ${{\bf{X}}_2}$ 
 \[{\mathbf{\Sigma  = }}\left( {\begin{array}{*{20}{c}}
  {{{\mathbf{\Sigma }}_{{\mathbf{11}}}}}&{{{\mathbf{\Sigma }}_{{\mathbf{12}}}}} \\ 
  {{{\mathbf{\Sigma }}_{{\mathbf{21}}}}}&{{{\mathbf{\Sigma }}_{{\mathbf{22}}}}} 
\end{array}} \right)\]
The  canonical vectors $ \textbf{u} $ and $ \textbf{v}$ (or factors) are respectively given by the norm 1 eigenvectors of  ${\bf{\Sigma }}_{11}^{ - 1}{{\bf{\Sigma }}_{12}}{\bf{\Sigma }}_{22}^{ - 1}{{\bf{\Sigma }}_{21}}$ and ${\bf{\Sigma }}_{22}^{ - 1}{{\bf{\Sigma }}_{21}}{\bf{\Sigma }}_{11}^{ - 1}{{\bf{\Sigma }}_{12}}$. 
They define the linear combinations of the columns of ${{\bf{X}}_1}$ and ${{\bf{X}}_2}$   yielding the canonical variates ${{\bf{X}}_1}{\bf{u}}$ and ${{\bf{X}}_2}{\bf{v}}$.

CCA of indicator matrices is related to one of the first emergence of CA as a way of scaling optimally the categories of two nominal variables (Fisher, 1940) \textit{cf.} de Leeuw (1973).

Let ${{\bf{X}}_1}$ and ${{\bf{X}}_2}$  be the data tables with $n$ rows and respectively $I$ and $J$ columns which are the indicator (binary) variables of the categories of two categorical variables.
 \[{{\bf{X}}_1} = \left( {\begin{array}{*{20}{c}}
1&0&0&0\\
1&0&0&0\\
.&.&.&.\\
0&1&0&0
\end{array}} \right){\rm{   }}{{\bf{X}}_2} = \left( {\begin{array}{*{20}{c}}
0\\
0\\
.\\
1
\end{array}{\rm{ }}\begin{array}{*{20}{c}}
1\\
0\\
.\\
0
\end{array}{\rm{ }}\begin{array}{*{20}{c}}
0\\
1\\
.\\
0
\end{array}} \right)\]
If we cross-classify these categorical variables, the contingency table is $${\bf{N = }}{{\bf{X}}_1}{\bf{X}}_2^{\bf{'}} = {{\bf{\Sigma }}_{12}}$$ 

Note that in this particular case and forgetting that the indicator variables are not centered:  ${{{\bf{X'}}}_1}{{\bf{X}}_1} = n{{\bf{D}}_r} = {{\bf{\Sigma }}_{11}}$ and ${{{\bf{X'}}}_2}{{\bf{X}}_2} = n{{\bf{D}}_c} = {{\bf{\Sigma }}_{22}}$ are diagonal matrices.

With some simple algebra it can be easily shown that canonical correlation analysis between ${{\bf{X}}_1}$ and ${{\bf{X}}_2}$ will give the same results as the CA of $ \textbf{N} $. The eigenvalues of CA are the squared canonical correlations. Pairs of canonical variates are linear combinations of resp. the $I$ and $J$ variables, the coefficients of these combinations being the standardized coordinates ${\bf{a}}/\sqrt \lambda  $  and  ${\bf{b}}/\sqrt \lambda  $. 
Centering  ${{\bf{X}}_1}$ and ${{\bf{X}}_2}$ avoids the trivial solution $\lambda  = 1$ .
It is important to note that with the CCA approach one does not need to introduce chi-square distances nor profile weights.

\subsection{CA as a MCA with \textit{p} = 2 variables}
Let ${\bf{X}} = \left( {{{\bf{X}}_1} \vdots {{\bf{X}}_2}} \right)$  be the concatenation of ${{\bf{X}}_1}$ and ${{\bf{X}}_2}$.
Then, apart from scaling constants, performing a formal CA on $  \textbf{X}$ is equivalent to perform a CA on $ \textbf{N} $. Since the rank of $  \textbf{X}$ is larger than the rank of $ \textbf{N} $, one should keep only the $min(I-1, J-1)$ largest eigenvalues.
In general terms, when analyzing $p> 2$ categorical variables MCA is nothing else than CA applied to the “complete disjunctive binary table” ${\bf{X}} = \left( {{{\bf{X}}_1} \vdots {{\bf{X}}_2} \vdots  \cdots  \vdots {{\bf{X}}_p}} \right)$  or to the so-called Burt's table $\textbf{B}$ grouping all contingency tables (Lebart \& Saporta, 2014): 
\[{\bf{B = X'X = }}\left( {\begin{array}{*{20}{c}}
{{\bf{X}}_{\bf{1}}^{\bf{'}}{\bf{X}}_{\bf{1}}^{}}&{{\bf{X}}_{\bf{1}}^{\bf{'}}{{\bf{X}}_{\bf{2}}}}&{..}&{{\bf{X}}_{\bf{1}}^{\bf{'}}{{\bf{X}}_{\bf{p}}}}\\
{{\bf{X}}_{\bf{2}}^{\bf{'}}{{\bf{X}}_{\bf{1}}}}&{{\bf{X}}_{\bf{2}}^{\bf{'}}{\bf{X}}_{\bf{2}}^{}}&{..}&{{\bf{X}}_{\bf{2}}^{\bf{'}}{{\bf{X}}_{\bf{p}}}}\\
{..}&{..}&{..}&{..}\\
{{\bf{X}}_{\bf{p}}^{\bf{'}}{\bf{X}}_{\bf{1}}^{}}&{{\bf{X}}_{\bf{p}}^{\bf{'}}{\bf{X}}_{\bf{2}}^{}}&{..}&{{\bf{X}}_{\bf{p}}^{\bf{'}}{\bf{X}}_{\bf{p}}^{}}
\end{array}} \right)\]

\section{Useful sparse versions of some multivariate \linebreak methods}
\subsection{Sparse PCA  }
Due to its two main properties : maximal variance and orthogonality of the components,  PCA is widely and successfully used in many applications. However, when the number of variables is large the interpretation of the principal components may become difficult, since each component is a linear combination of all variables. 
When the number of variables is much larger than the number of observations, an other less known problem occurs: \textit{inconsistency} which means that the sample principal components may not converge towards the population principal components (Hall \textit{et al}., 2005).
Sparse PCA where components would be linear combinations of a small number of variables is thus appealing, but there is no unique definition of sparse PCA. 
There is a trade-off between sparsity and variance: a large number of zeros makes the interpretation easier but leads to a poor approximation, \textit{cf} later part 3.5.1
  
Over the past fifteen years, a large number of variants of sparse PCA has been proposed. In their review paper Shen Ning-min \& Li Jing (2015) count about twenty algorithms that they divide into 3 classes related to different viewpoints on PCA: 
\begin{itemize}
\item data-variance-maximization
\item minimal-reconstruction-error 
\item probabilistic modeling viewpoint
\end{itemize}
The sparse PCA variants are mostly based on the first two classes.
 
a.	Since the principal components are the linear combination of the input variables with maximal variance, a first set of methods consist in imposing  sparsity constraints to the maximisation problem :   $ \max({\bf{v'\Sigma v}})$. 

b.	The second viewpoint  starts from finding a low rank $k$ approximation of $\textbf{X} $ denoted ${{\bf{\hat X}}_k}$ that minimizes the total squared reconstruction error ${\left\| {{\bf{X}} - {{{\bf{\hat X}}}_k}} \right\|^2}$ which is the Frobenius norm of the difference between both matrices.
The Eckart-Young theorem states that ${{\bf{\hat X}}_k}$ is obtained by the truncated singular value decomposition (SVD) of $\textbf{X} $ of order $k $ :

$${{{\bf{\hat X}}}_k} = \sum\limits_{j = 1}^k {{\alpha _j}{{\bf{u}}_j}{{\bf{v}}_j}'} $$

Sparse SVD has been proposed with sparsity constraints on both $\textbf{u}$ and $\textbf{v}$, or $\textbf{v}$ alone. 

\subsubsection{PCA with additional  L1  constraints}
   
\qquad a.	Using  L1  constraints is one of the simplest ways to obtain sparse weights. SCoTLASS (Joliffe \textit{et al}., 2003) which stands for Simplified Component Technique for Least Absolute Shrinkage and Selection is considered as the first true algorithmic method to achieve sparsity. It consists in adding the extra constraint ${\left\| {\bf{v}} \right\|_1} \leqslant \tau $ where ${\left\| {\bf{v}} \right\|_1} = \sum\limits_{j = 1}^p {\left| {{v_j}} \right|} $ to the classical maximization of the variance: $\max {\bf{v'\Sigma v}}{\text{  with }}{\left\| {\bf{v}} \right\|^2} = {\bf{v'v}} = 1{\text{ }}$. It is necessary that $1 < \tau  < \sqrt p {\text{ }}$ since if $\tau  \geqslant \sqrt p $ we have usual PCA, and if $\tau  = 1$ there is only one non zero weight.
SCoTLASS is a non convex and computationally costly algorithm which generally prevents to test many values of the sparsification (or tuning) parameter $\tau$.

b.	Zou \& Hastie (2006) proposed SPCA, a more efficient algorithm based upon a ridge regression property of PCA:  they observe that each PC being a linear combination of the $p$ variables, its weights can be recovered by a ridge regression of the principal component onto the $p$ variables:
a principal  component $\textbf{z}$ and the associated weight vector ${\bf{\beta }}$  are the solution of 

\[{{\bf{\hat \beta }}_{ridge}} = \mathop {{\rm{arg \min}}}\limits_{\rm{\beta }} \left[ {{{\left\| {{\bf{z - X\beta }}} \right\|}^2} + \lambda {{\left\| {\bf{\beta }} \right\|}^2}} \right]\]

 Sparse weights are produced by adding a L1  constraint:

\begin{equation}
{\bf{\hat \beta }} = \mathop {\arg \min }\limits_\beta  \left[ {{{\left\| {{\bf{z - X\beta }}} \right\|}^{\bf{2}}}{\bf{ + }}\lambda {{\left\| {\bf{\beta }} \right\|}^{\bf{2}}} + {\lambda _1}{{\left\| {\bf{\beta }} \right\|}_1}} \right]
\end{equation}

${\lambda _1}$ is the sparsity parameter, the larger it is, the more weights are null.
 The solution in obtained by iterating elastic net and SVD.
In their applications, it was found that SPCA account for a larger amount of variance   with a much sparser structure than the SCoTLASS.

\subsubsection{Prespecification of the number of zeros  }
Prespecifying the number of zeros for the first component it is equivalent to use a  L0  norm constraint.  Adachi \& Trendafilov (2015) proposed a method  called unpenalized sparse loading PCA (USLPCA) in which the total number of nonzero loadings (the cardinality of the loading matrix) for a set of $k$ components is pre-specified, without using penalty functions. In their approach, note that they sparsify the loading matrix $\textbf{A}$ instead of the weight matrix $\textbf{V}$. See later part 3.5.4  

\subsection{Sparse MCA}
In MCA since the columns of $\textbf{ X}$ represent categories of nominal variables, Bernard \textit{et al}., (2012) proposed to globally select the variables, \textit{ie} blocks of $\textbf{ X}$  and not separate columns or categories. Their penalty approach relies upon a modification of equation (8) similar to the sparse-group lasso of Simon \textit{et al.} (2013)
\begin{equation}
 {{\bf{\hat \beta }}_{GL}} = \mathop {{\rm{arg \min}}}\limits_{\bf{\beta }} \left[ {{{\left\| {{\bf{z}} - \sum\limits_{j = 1}^J {{{\bf{X}}_j}} {{\bf{\beta }}_j}} \right\|}^2} + \lambda \sum\limits_{j = 1}^J {\sqrt {{p_j}} } {{\left\| {{{\bf{\beta }}_j}} \right\|}_1}} \right]
\end{equation}

The algorithm consists in alternating sparse-group lasso and SVD until convergence.

Mori \textit{et al}. (2016) proposed a different method based upon the pre-specifying sparsity approach of USLPCA.

It is important to note that one cannot derive a sparse CA from a sparse MCA: CA is equivalent to a MCA with only 2 blocks, and sparse MCA selects entire blocks of indicator variables. Selecting a block, \textit{ie} discarding one of the 2 categorical variables, does not make sense here.

\subsection{Sparse canonical correlation analysis}
As noticed in section 2.4, a canonical correlation analysis between indicator matrices  ${{\bf{X}}_1}$ and ${{\bf{X}}_2}$ corresponding to the variables categories will give the same results as the CA of the contingency matrix $\textbf{N}$.  However, in high dimensional data, when the variables outnumber the sample size or when the variables are highly correlated classical CCA is no longer appropriate. Like in high dimensional PCA, the main drawbacks are the unstability of the estimates and the lack of interpretability of the combinations based on a large number of original variables.  
In the CA context this may happen in case of large number of categories for one or both variables under study. For example in text mining one variable may represent categories of documents or authors and the other variable columns correspond to the terms of these documents. 
To overcome these traditional CCA limitations, sparse versions of CCA have been developped by several authors. They all improve the interpretability of canonical variables by restricting the linear combinations to small subsets of variables.

Wilms \& Croux (2015) present a short review of the main existing methods before proposing their own sparse CCA method. As for sparse PCA, the sparse CCA methods can be grouped into two main families : one based on SVD and one in the regression framework. 
Parkhomenko \textit{et al}. (2009) consider singular value decomposition to derive sparse singular vectors through an efficient iterative algorithm that alternately approximates the left and right singular vectors of the SVD using iterative soft-thresholding for feature selection.  Their sparse CCA (SCCA) method seeks sparsity in both sets of variables simultaneously. It incorporates variable selection and produces linear combinations of small subsets of variables from each group of measurements with maximal correlation. 

A similar approach was taken by Witten \textit{et al}. (2009) who apply a penalized matrix decomposition to the cross-product matrix ${\bf{X}}_{\bf{1}}^{\bf{'}}{{\bf{X}}_{\bf{2}}}$

In the general case, a limitation of these approaches is that they require  the variables within each of the two datasets to be uncorrelated  to guaranteed sparsity in the canonical vectors. But, in the particular case of correspondence analysis, where the variables are the categories indicators, the previous limitation is naturally satisfied since within each set, the indicator variables are orthogonal and consequently the matrices   ${\bf{X}}_{\bf{1}}^{\bf{'}}{{\bf{X}}_{\bf{1}}}$ and   ${\bf{X}}_{\bf{2}}^{\bf{'}}{{\bf{X}}_{\bf{2}}}$ are diagonal and regular.

Wilms \& Croux (2015) consider the CCA problem from a predictive point of view and reformulate it into a regression framework. They induce sparsity in the canonical vectors by combining an alternating penalized regression approach with a lasso penalty.

\subsection{Sparse SVD }
The Penalized Matrix Decomposition (PMD) approach by Witten \textit{et al}.(2009) solves the following optimization problem for the first pair of left and right singular vectors: 
\begin{equation}
\max {\bf{u}}'{\bf{Xv}}{\text{ with }}{\left\| {\bf{u}} \right\|^2} = {\left\| {\bf{v}} \right\|^2} = 1{\text{ and }}{P_1}({\bf{u}}) \leqslant {c_1}{\text{  }}{P_2}({\bf{v}}) \leqslant {c_2}
\end{equation}
where $P_1$ and $P_2$ are convex penalty functions such as, \textit{e.g.}, the LASSO ${P_1}({\bf{u}}) = \sum\limits_{i = 1}^n {\left| {{u_i}} \right|} $ and ${P_2}({\bf{v}}) = \sum\limits_{j = 1}^p {\left| {{v_j}} \right|} $ or the fused LASSO constraints with $c_1$ and $c_2$  being positive constants.
Note that  $\max {\bf{u}}'{\bf{Xv}}{\text{ }}$ is equivalent to  $\min{\left\| {{\bf{X}} - {\bf{uv}}'} \right\|^2}$ with the same constraints.

Following Witten \textit{et al}. $\sum\limits_{i = 1}^n {\left| {{u_i}} \right|} $  and $\sum\limits_{j = 1}^p {\left| {{v_j}} \right|} $  will be denoted \textit{sumabsu} and \textit{sumabsv}  respectively.\textit{ sumabsu} (resp \textit{sumabv}) must be between 1 and the square root of the number of rows  (resp. columns)  of $\textbf{X}$. The smaller it is, the sparser $\textbf{u}$ (resp $\textbf{v})$ will be.  The pseudo singular value  $\alpha$  is then equal to ${\bf{u}}'{\bf{Xv}}$ .

When \textit{sumabsu} is large, \textit{ie} when there is no penalty on $\textbf{u}$, PMD is equivalent to the sparse PCA of $\textbf{X}$ with the  SCoTLASS criterium but with a different algorithm.
sPCA-rSVD which stands for sparse PCA via a one- sided regularized SVD (Shen \& Huang, 2008) is also a special case of PMD since its criterium is $\mathop {\min }\limits_{{\bf{u}}{\bf{,v}}} \left[ {{{\left\| {{\bf{X}} - {\bf{uv}}'} \right\|}^2} + P_\lambda ^{}({\bf{v}})} \right]$ with $\left\| {\bf{u}} \right\| = 1$  and ${P_\lambda ^{}}$ a penalty function which is usually the lasso or L1 penalty. 

If one looks for a similar degree of sparsity on $\textbf{u}$ and $\textbf{v}$, Witten \textit{et al}. suggest to use a unique parameter denoted \textit{sumabs} such that: \\
$sumabsu = \sqrt n$   $sumabs$ and $sumabsv = \sqrt p$  $sumabs$ .

\subsection{General issues}
In addition to the usual question “ how many principal components should we keep? ”, sparse exploratory methods raise specific issues.
\subsubsection{Tuning sparsity parameters }
Since sparse PCA is an unsupervised technique, there is no well-established criterium to choose the tuning parameters, \textit{ie} the degree of sparsity. As Trendafilov \textit{et al}. (2017) wrote: \textit{“The sparser solutions give worsen fit to the data. Thus, the big problem of any sparse technique is to compromise between sparseness and goodness of fit}.” Using different degrees of sparsity for each component increases the difficulty.

\begin{itemize}
\item 	Empirical approaches have been proposed. Some are based upon scree-plots where one plots the amount of explained variance, against the sparsity parameter or against the number of zeros. One then tries to detect visually a breakdown in the slope of the curve. Plotting the coefficients against the tuning parameter in a similar way as the regularization paths of the Lasso gives also important information and help selecting the parameter value.
\item 	Trendafilov \textit{et al}. (2017) proposed to maximize the following index of sparsity:
\begin{equation}
   IS(\tau ) = \frac{{{\text{original fit}}}}{{{\text{fit for }}\tau }}{\left( {\frac{{{\# _0}}}{{pr}}} \right)^2} 
  \end{equation}  
  Where  $\tau$ is defined by the ScoTLASS constraint ${\left\| {\bf{u}} \right\|_1} \leqslant \tau $ . The fit is measured by the total variance accounted for $r$ components, $\#0$ is the number of zeros among all $pr$ loadings or weights. 
\item	Cross validation has been proposed by Witten \textit{et al}. (2008). It consists in deleting 10\% of the elements of the data matrix $ \textbf{X} $, then imputing the missing data by sparse SVD, repeat it ten times for several values of the sparsity parameter and choose the best value in terms of mean square error. 
\item 	Following Zou \textit{et al}. (2007), Shen \textit{et al}. (2013) proposed to select for each component the sparsity parameter which minimizes the following BIC criterium : 
\begin{equation}
BIC(\tau ) = \frac{{{{\left\| {{\bf{X}} - {\bf{\hat X}}} \right\|}^2}}}{{np{{\hat \sigma }^2}}} + \frac{{\ln (np)}}{{np}}df(\tau)
\end{equation}
where ${\hat \sigma ^2}$  is the ordinary-least squares estimate of the error variance and $df(\tau )$ the number of non zero weights (or loadings, see later) for the threshold parameter.
\end{itemize}

\subsubsection{More than one component: deflation and lost properties}
The search for more than one component is not obvious. Once the first sparse principal component is obtained, higher order solutions are computed by some kind of deflation. 
Since the comprehensive survey of Mackey (2008), there has been many papers dealing with this topic. In standard PCA, all deflations methods provide the same solutions, but it is not the case with sparse PCA. Mackey points out that Hotelling's deflation does not in general preserve positive-semidefiniteness when applied to a noneigenvector and suggests several alternative techniques. 

The simplest and most appealing one (at least for us) is  the \textit{projected deflation} which consists in projecting the data table $\textbf{X}$  onto the orthocomplement of the space spanned by the sparse pseudo eigenvector $\textbf{v}$.  $\textbf{X}$ becomes ${\bf{X}}\left( {{\bf{I}} - {\bf{vv'}}} \right)$.
It comes down to pre and post multiplying the variance-covariance matrix by the projection matrix $\left( {{\bf{I}} - {\bf{vv'}}} \right)$: $$\left( {{\bf{I}} - {\bf{vv'}}} \right){\bf{\Sigma }}\left( {{\bf{I}} - {\bf{vv'}}} \right)$$
In standard PCA, the principal components are uncorrelated and their loadings are orthogonal. These properties are lost in sparse PCA and similar techniques: one cannot have both orthogonality for loadings and for components. In SCoTLASS the successive “principal” weights vectors are forced to be orthogonal, but successive components are no longer uncorrelated. 

In SPCA,   sPCA-rSVD and PMD even the weights may not be orthogonal. 
In PMD, the second pseudo-singular triplet is estimated by solving the same optimization problem where  $\textbf{X}$  is replaced by the deflated matrix ${\bf{X}} - \alpha {\bf{uv'}}$ obtained by substracting the rank one matrix $\alpha {\bf{uv'}}$ defined by the sparse left and right eigenvectors, but the solution is not orthogonal to this rank one matrix.

Rather than the previous case (close to Hotelling’s deflation) we propose the following one inspired by the \textit{projected deflation} : we replace $\textbf{X}$  by $$\left( {{\bf{I}} - {\bf{uu'}}} \right){\bf{X}}\left( {{\bf{I}} - {\bf{vv'}}} \right)$$ We will call this method \textit{projected PMD} or pPMD.

In a recent paper, Guillemot \textit{et al}. (2019) presented CSVD (Constrained Singular Value Decomposition) a new algorithm and an associated \texttt{R} package that implements orthogonality constraints on successive sparsified singular vectors as the projection onto the intersection of the convex sets expressing the constraints. 

PMD, pPMD and CSVD provide the same solution for the first dimension, and we will show on real data sets, that pPMD provides near orthogonal weights and components vectors.

It may also happen that components obtained by some sparse method are not ordered according to the magnitudes of their variances, corrected or not. It is a consequence of the non orthogonality of components.  This annoying point is apparently neglected in many publications that focus only on the first-rank solution.

\subsubsection{Proportion of variance explained}
When the sparse principal weights are not orthogonal and/or the sparse principal components are correlated, the variances of sparse components are not additive. 
Several solutions have been proposed among which we keep the following one proposed by  Shen \& Huang (2008):  the proportion of variance explained by the first $k$ sparse principal components is defined as
$$\frac{{Trace\left( {{\bf{\tilde X}}_{\bf{k}}^{\bf{'}}{{{\bf{\tilde X}}}_{\bf{k}}}} \right)}}{{Trace\left( {{\bf{X'X}}} \right)}}$$
where ${{\bf{\tilde X}}_{\bf{k}}} = {\bf{X}}{{\bf{V}}_{\bf{k}}}{\left( {{{\bf{V}}_{\bf{k}}}{\bf{V}}_{\bf{k}}^{\bf{'}}} \right)^{ - 1}}{\bf{V}}_{\bf{k}}^{\bf{'}}$ is the projection of $\textbf{X}$ onto the $k$-dimensional subspace spanned by the $k$ principal sparse vectors. ${{\bf{V}}_{\bf{k}}}$  is the matrix that has the first $k$ sparse weights vectors as its columns.

\subsubsection{Sparse weights or sparse loadings?}
One should make a clear distinction between the weights and the loadings. The weights vector $\textbf{v}$ gives the coefficients of the original variables that provide a new variable or component by the linear combination $\textbf{z} =\textbf{Xv}$.
If we use $k$ components the weights matrix $\textbf{V}$ ($p,k$) is such that the component (or score) matrix is $\textbf{Z}=\textbf{XV}$ ($n,k$). 
The loadings are the covariances between components and original variables, collected in the matrix $\textbf{A}$: 
$${\bf{A}} = \frac{1}{n}{\bf{X'Z}} = \frac{1}{n}{\bf{X'XV}} = {\bf{\Sigma V}}$$
In usual PCA, since the $ \textbf{v} $’s are norm-1 eigenvectors of the variance-covariance matrix ${\bf{\Sigma }}$, the loadings are proportional to the weights: ${\bf{A}} = {{\bf{V}{\bf{D}}_\lambda }}$. 

Most sparse PCA techniques look for a sparse weights matrix $ \textbf{V} $ and thus $\textbf{A}  $ is not sparse. Looking for sparse loadings was the aim of rotations in factor analysis. Only a few sparse PCA algorithms like USLPCA from Adachi \& Trendafilov (2015) are looking for sparse loadings.  

\section{ Sparse Correspondence Analysis}
Correspondence analysis is at the crossroads of several methods, of which it is a special case. Since sparse versions of these different methods have been developed, our proposition is to use them wisely to develop a sparse version of correspondence analysis. 
    
However, the diversity of sparse approaches implies a similar diversity for the sparsification of correspondence analysis. Many algorithms and packages also exist, sometimes of imperfect quality. So choices had to be done. We will therefore limit ourselves for the rest of this paper to two cases: 
\begin{itemize}
\item 	The first one (doubly sparse) looks for underlying dimensions that are explained by sparse combinations of both rows and columns. Sparse SVD provides this solution.
\item 	The second one (column sparse only) is an analysis where the weights of the columns are sparsified but not those of the rows. It is adapted to the case where we try to reduce the number of columns, but not the number of rows, to describe each dimension. 
\end{itemize}
\subsection{Doubly sparse CA}
This analysis is identical to a sparse SVD of ${\bf{Z}} = {\bf{D}}_r^{ - 1/2}\left( {{\bf{P}} - {\bf{rc}}'} \right){\bf{D}}_c^{ - 1/2}$ 
Rows and columns coordinates $  \textbf{a}$ and $ \textbf{b} $ are obtained by mutiplying the row and column profiles matrices by the left and right sparse “principal” $ \textbf{u} $  and $\textbf{v}  $ and normalizing the results in order to have a variance equal to $\lambda  = {\alpha ^2}$

\begin{equation}
\begin{array}{l}
{\bf{a}} \propto {\bf{D}}_r^{{\bf{ - 1}}}{\bf{Pv}}{\rm{  }}\quad with \quad{\rm{ }}{\bf{a'}}{{\bf{D}}_{\rm{r}}}{\bf{a}}{\rm{ = }}\lambda \\
{\bf{b}} \propto {\bf{D}}_c^{{\bf{ - 1}}}{\bf{P}}'{\bf{u}}{\rm{ }}\quad with \quad{\rm{  }}{\bf{b'D}}{}_{\rm{c}}{\bf{b}}{\rm{ = }}\lambda 
\end{array}
\end{equation}

Pseudo-barycentric formulas are no longer true: unlike standard CA, $  \textbf{v}$  is not proportional to $ \textbf{b} $ and $  \textbf{u}$   is not proportional to $  \textbf{a}$  : in other words, \textit{the weights differ from the coordinates}. The usual table of coordinates and contributions should be replaced by the table of weights and contributions which highlights the zeros. An important point is that the maps should use the coordinates \textbf{a} and \textbf{b} and not the weights \textbf{u} and \textbf{v}:  otherwise there is a risk of having too many points stuck to the axes, those with zero weights,  and thus uninterpretable maps as shown in Figure 1.

\begin{figure}[hbtp]
\centering
\includegraphics[width=9cm]{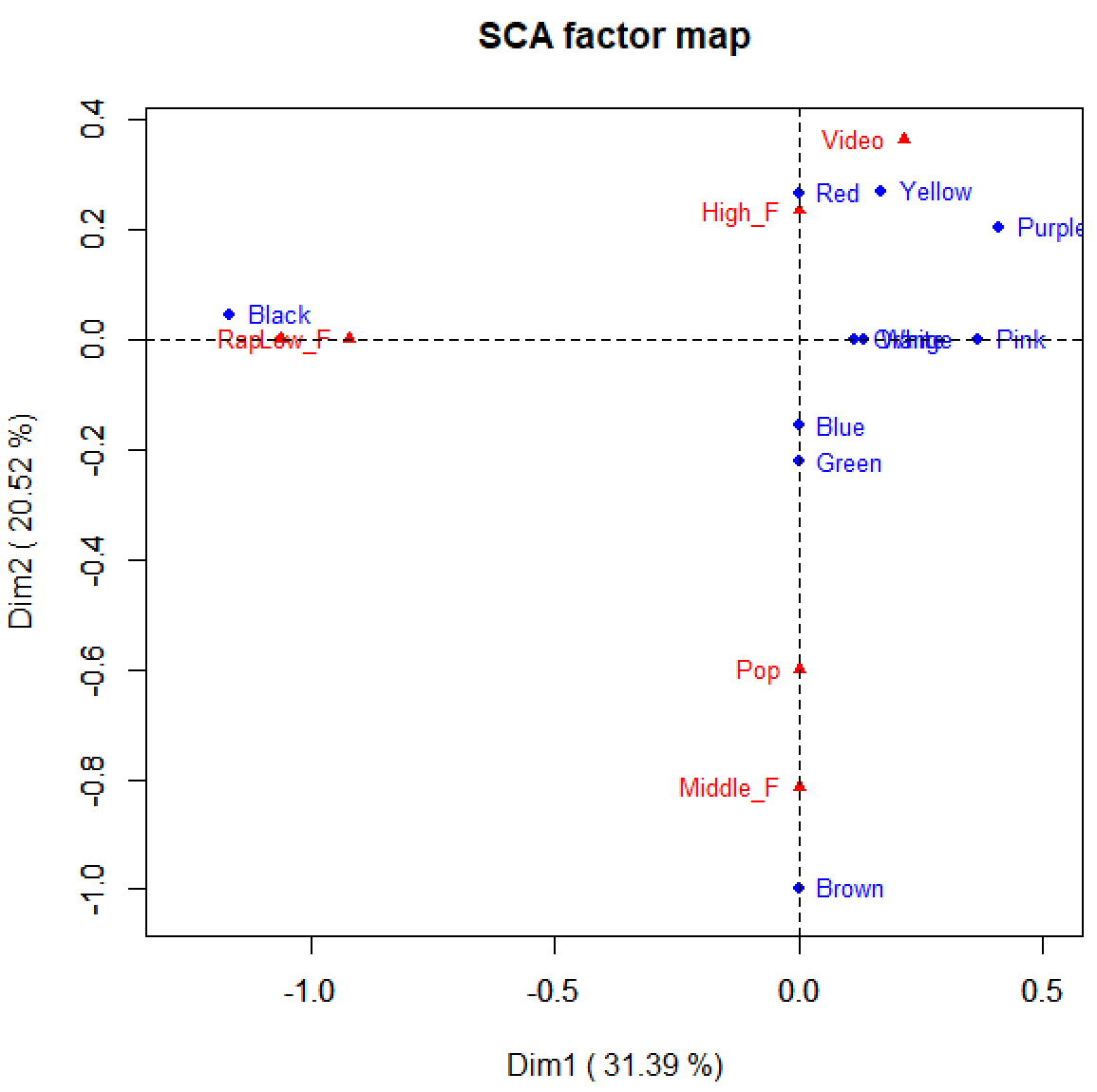}
\caption{An example of a weight map}
\end{figure}
\pagebreak 
\subsection{Column sparse CA}
There are two ways for introducing this case:\\

a. Start from doubly sparse CA and relax the constraint on $\sum\limits_{i = 1}^I {\left| {{u_i}} \right|} $\\

b. Perform a sparse PCA of the row profiles table ${\bf{D}}_r^{ - 1}{\bf{P}}$ with row weights  ${{\bf{D}}_r}$ and chi-square metric ${\bf{D}}_c^{ - 1}$.
 \\

Let us detail the second way: using notations of part 2.2, let $\textbf{v}$ be the vector containing the sparse normalized weights of the columns obtained by a s-PCA method and scaled such as ${\bf{v'}}{{\bf{D}}_{\bf{c}}}{\bf{v = }}1$

The vector of row coordinates  $\textbf{a}$ is proportional to ${\bf{D}}_r^{ - 1}{\bf{Pv}}$  and scaled such as the weighted variance of the row coordinates be equal to the pseudo eigenvalue:  ${\bf{a'}}{{\bf{D}}_r}{\bf{a}} = \lambda $ 

Row profiles may be displayed as usual in two dimensional plots, but displaying simultaneously rows and columns profiles raises a difficulty similar to a simultaneous display of units and variables in PCA.

By construction ${\bf{a}} \propto {\bf{D}}_r^{ - 1}{\bf{Pv}}$ is a transition or pseudo-barycentric equation for the rows: along each axis, the row points are pseudo barycenters of the column weights, but the symmetric property is not preserved: column weights are not pseudo barycenters of row coordinates.  Nevertheless the previous property is not appropriate to display column profiles as pointed in the previous part, since many columns weights should be equal to zero.

We will use the following solution: row profiles are displayed by means of the pseudo principal components $ \textbf{a} $ “as usual”. Column categories are then displayed as barycenters of the row points like in CA $${\bf{b = D}}_c^{ - 1}{\bf{P'a}}$$ The differences with CA is that the $  \textbf{b}$’s are not principal components of the column profiles and that the points representing row profiles are not pseudo barycenters of the points representing the column profiles.
Like in asymmetric maps the column points are less spread than row points. If one wants both clouds equally spread,  $ \textbf{b} $ has to be scaled in the same way as $ \textbf{a} $: ${\bf{b}}'{{\bf{D}}_c}{\bf{b = }}\lambda $
\subsection{General approach}
In order to perform a sparse CA, the following steps are necessary:

-	method choice: double or single sparse
 
-	tuning (sparsity) parameters choice

-	deflation

-	interpretation

In this paper we limit ourselves to the projected PMD or pPMD introduced in 3.5.2.  If only column sparsity is needed, like for sparse PCA, one has to relax the L1 constraint on $ \textbf{u} $.

The choice of the tuning parameters requires some skill, especially for large data sets where the automatic criteria may give solutions which are not sparse enough; in the latter case it is recommended to set the number of non zero coefficients in advance. 
It should be noted that among lost properties, the simultaneous barycentric equations (3) do not hold, since they are a characteristic property of standard CA.

\subsection{A toy example }

We apply sparse CA to the data of table 1 coming from Abdi \& Béra (2017): \textit{J}= 9 pieces of music were presented to 22 participants who were asked to associate one of \textit{I}=10 colors to each piece of music. 
Vectors ni. and r are the row totals and the row frequencies, vectors n.j and c are the column totals and the column frequencies.
\pagebreak 

\begin{center}
Table 1: The colours of music
\end{center}
\begin{scriptsize}
\begin{filecontents*}{Table1.csv}
Color Sound,Video,Jazz,Country,Rap,Pop,Opera,LowF,HighF,MiddleF,ni.,r
Red,4,2,4,4,1,2,2,4,1,24,0.121
Orange,3,4,2,2,1,1,0,3,2,18,0.091
Yellow,6,4,5,2,3,1,1,3,0,25,0.126
Green,2,0,5,1,3,3,3,1,5,23,0.116
Blue,2,5,0,1,4,1,2,1,3,19,0.096
Purple,3,3,1,0,0,3,0,2,1,13,0.066
White,0,0,0,0,1,4,1,5,3,14,0.071
Black,0,2,0,11,1,3,10,1,1,29,0.146
Pink,2,1,1,0,2,4,0,2,0,12,0.061
Brown,0,1,4,1,6,0,3,0,6,21,0.106
n.j,22,22,22,22,22,22,22,22,22,N=198,1
c,0.11,0.11,0.1I,0.11,0.11,0.11,0.11,0.11,0.11,1,
\end{filecontents*}
\csvautobooktabular{Table1.csv}
\end{scriptsize}

\bigskip

\subsubsection{Standard CA}
Figure 2  and Tables 2 and 3 give the results of a standard CA. The first two components account for about 65\% of the total inertia. 
\begin{center}
Table 2: Eigenvalues, Percentages and Cumulated percentages
\end{center}

\begin{filecontents*}{Table2.csv}
,1,2,3,4,5,6,7,8
Eigenvalue,0.288,0.193,0.138,0.072,0.034,0.017,0.003,0
Percentage ,38.6,25.9,18.5,9.7,4.5,2.3,0.4,0
Cumulated percentage,38.6,64.5,83,92.7,97.3,99.6,100,100
\end{filecontents*}
\csvautobooktabular{Table2.csv}

\bigskip 

\begin{figure}[hbtp]
\centering
\includegraphics[width=9cm]{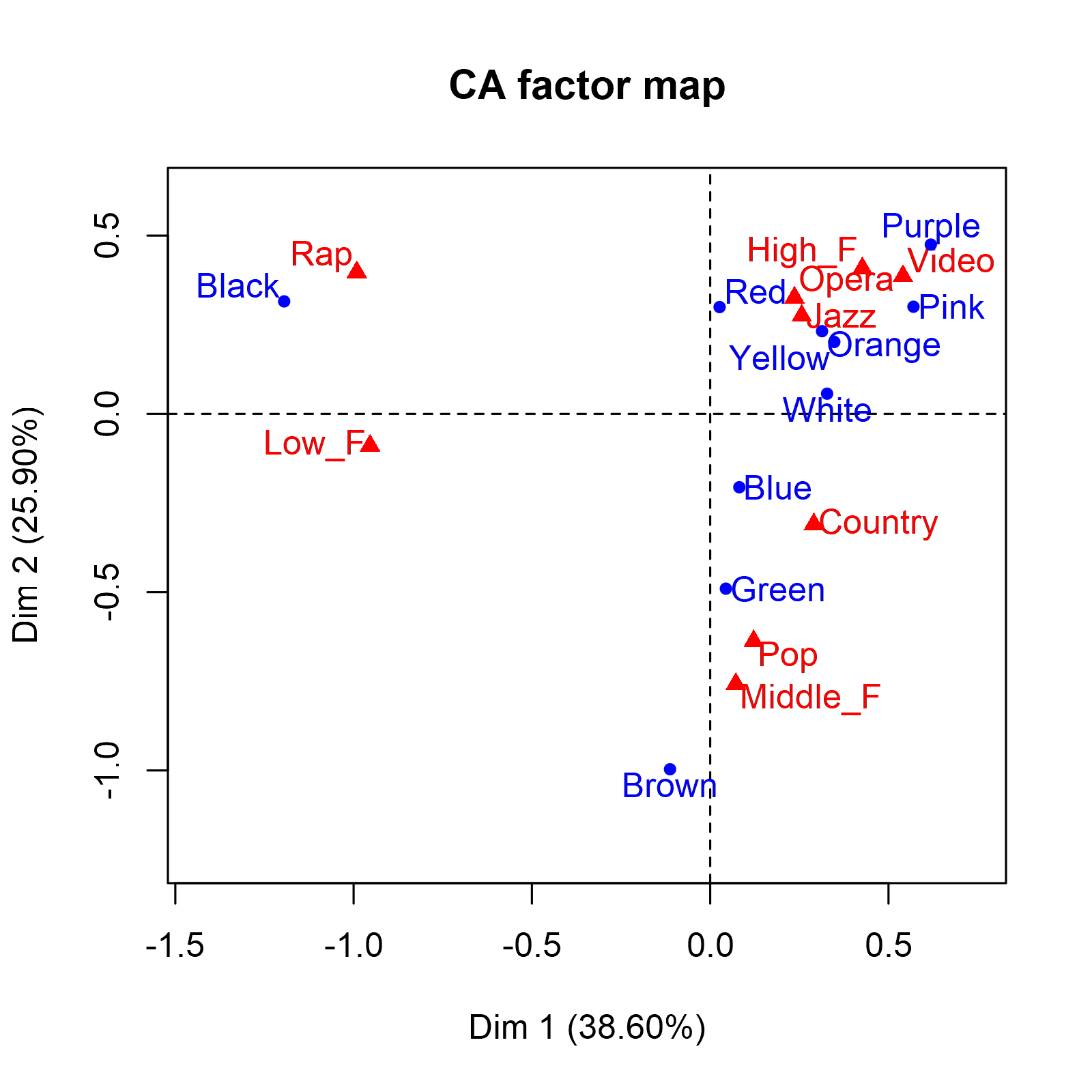}
\caption{CA map for colours of music }
\end{figure}

\pagebreak 
\begin{center}
Table 3: Rows and columns coordinates and contributions
\end{center}

\begin{center}
\begin{filecontents*}{Table3.csv}
,a1,a2,ctr1,ctr2,,,b1,b2,ctr1,ctr2
Red,0.026, 0.299,0,56,,Video,0.541,0.386,113,86
Orange,0.314, 0.232,31,25,,Jazz,0.257,0.275,25,44
Yellow,0.348, 0.202,53,27,,Country,0.291,-0.309,33,55
Green,0.044, -0.490,1,144,,Rap,-0.991,0.397,379,91
Blue,0.082, -0.206,2,21,,Pop,0.122,-0.637,6,234
Purple,0.619, 0.475,87,77,,Opera,0.236,0.326,22,61
White,0.328,  0.057,26,1,,LowF,-0.954,-0.089,351,5
Black,-1.195, 0.315,726,75,,HighF,0.427,0.408,70,96
Pink,0.570, 0.300,68,28,,MiddleF,0.072,-0.757,2,330
Brown,-0.113,-0.997,5,545,,,,,,
,,,,,,,,,,
total,,,1000,1000,,total,,,1000,1000
\end{filecontents*}
\csvautobooktabular{Table3.csv}
\end{center}
\bigskip

We reproduce here the comments of Abdi \& Béra:   \textit{The association with the low note reflects a standard association between pitch and color (high notes are perceived as bright and low notes as dark); by contrast, the association of rap music with the black color is likely to reflect a semantic association. The second component separates the colors brown and (to a lesser extent) green from the other colors (in particular purple) and that brown and green as associated with Pop and Middle F. On the other side of the second dimension, we find the color purple and a quartet of pieces of music (Video, High F, Opera, and Jazz) }
 
Since there is no reason to treat rows and columns in a different way, we opt now for a doubly sparse CA using projected penalized matrix decomposition (pPMD).

\subsubsection{Tuning the sparsity parameters for the first dimension}

In order to apply pPMD, we first have to choose the sparsity parameters  \textit{sumabsu} and  \textit{sumabsv}  which are equal respectively to $\sum\limits_{i = 1}^I {\left| {{u_i}} \right|} $ and $\sum\limits_{j = 1}^J {\left| {{v_j}} \right|} $. The smaller they are, the sparser $ \textbf{u} $ and $ \textbf{v} $ will be.\\

a.	Let us begin by looking for a unique parameter \textit{sumabs} if we want the same sparsity for rows an columns, such that $sumabsu = \sqrt I sumabs$ and $sumabsv = \sqrt J sumabs$. We have $1 < sumabsu < \sqrt I $ and $1 < sumabsv < \sqrt J $ therefore: $$\max \left( {\frac{1}{{\sqrt I }},\frac{1}{{\sqrt J }}} \right) < sumabs < 1$$
Figure 3, 4, 5 give the paths of rows and columns weights and of the proportion of zero weights when \textit{sumabs} varies.  We notice that some weights are never equal to zero, which corresponds to the fact that at least one row and one column should have a non zero weight. 

\begin{figure}[hbtp]
\centering
\includegraphics[width=9cm]{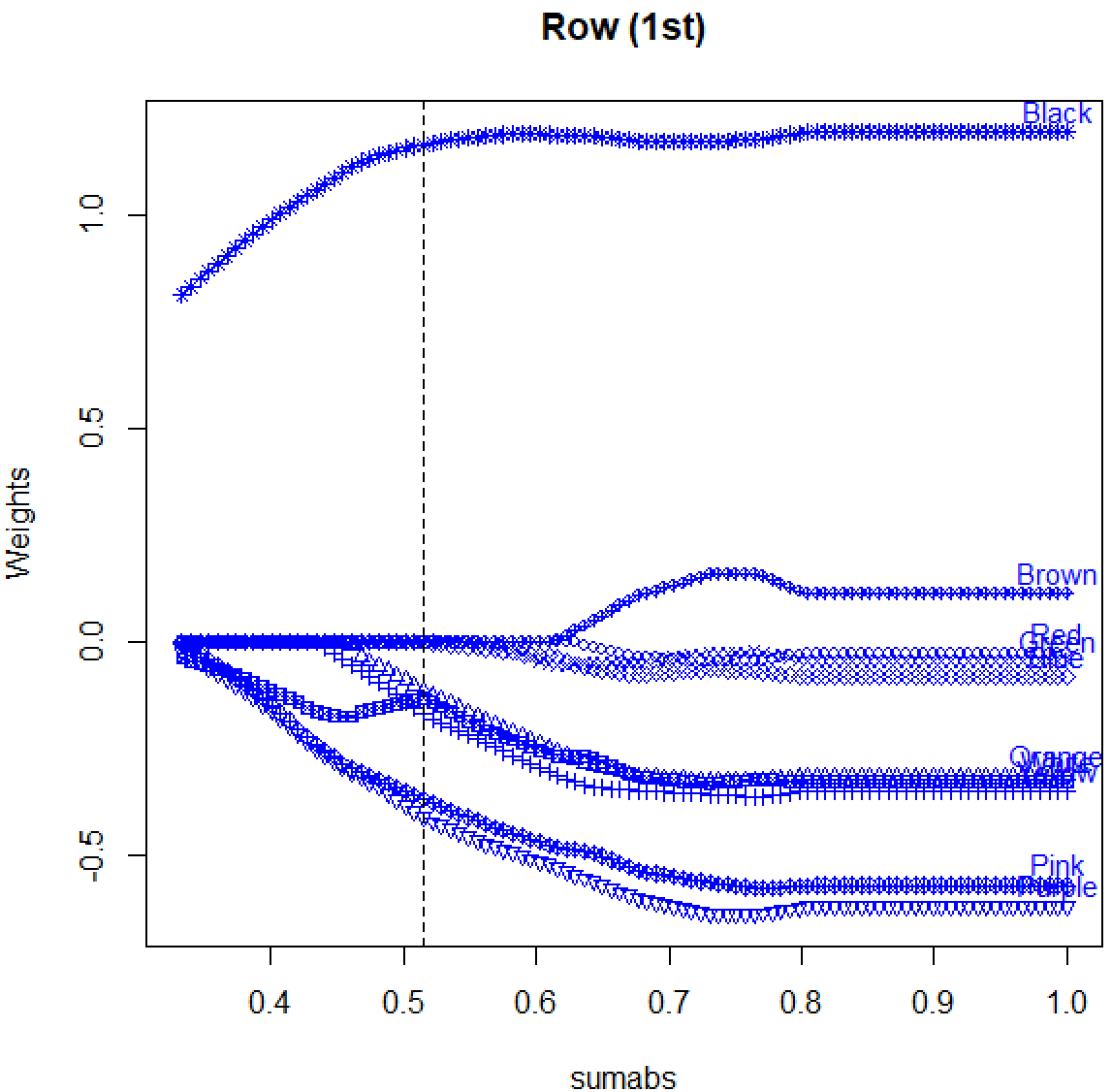}
\caption{First dimension row weights as a function of \textit{sumabs}. Dashed vertical line at optimal IS}
\end{figure}

\begin{figure}[hbtp]
\centering
\includegraphics[width=9cm]{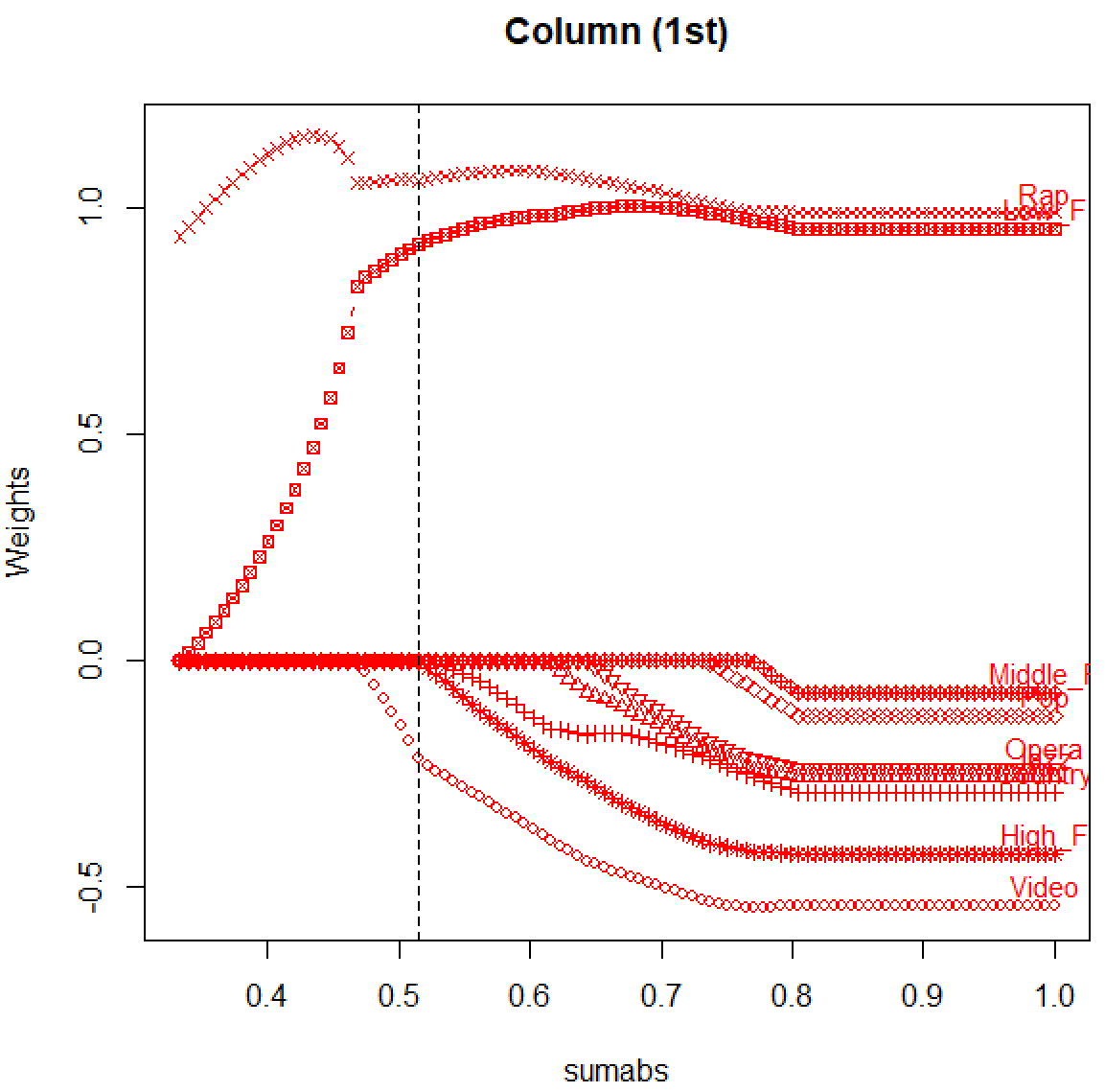}
\caption{First dimension column weights as a function of \textit{sumabs}. Dashed vertical line at optimal IS}
\end{figure}

\begin{figure}[hbtp]
\centering
\includegraphics[width=9cm]{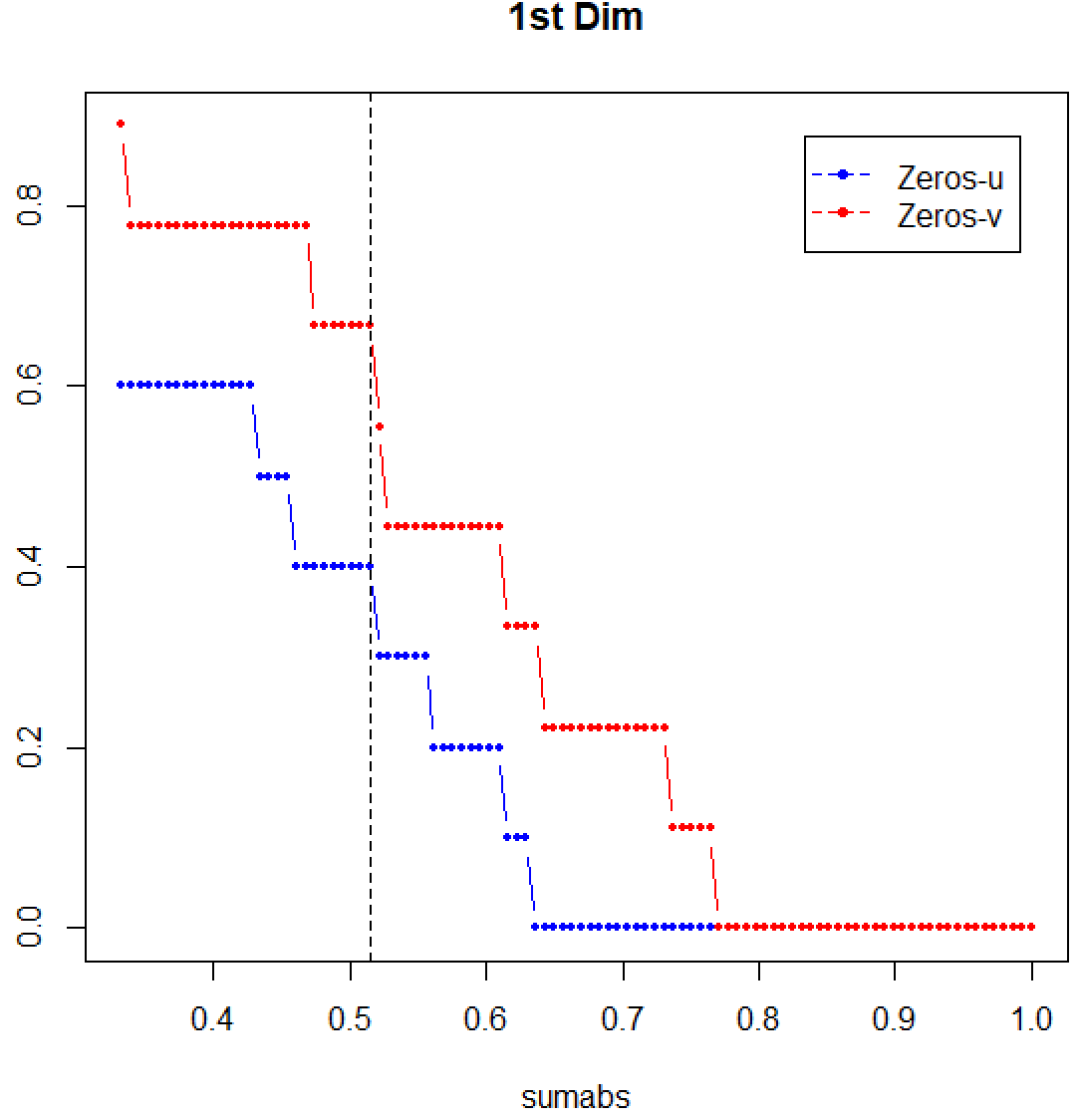}
\caption{Proportion of zero weights. Dashed vertical line at optimal IS}
\end{figure}

We will try now to use BIC, IS and CV criterion in order to select the best value of \textit{sumabs}. 
Figure 6 plots BIC as a function of \textit{sumabs}. The optimal value (minimal) of the BIC is obtained for \textit{sumabs}= 0.47  with 6 non zero row weights and 2 non zero column weights. It gives  \textit{sumabsu}=1.41 and \textit{sumabsv}=1.49 while the optimal value (maximal) of IS (Figure 7) is obtained for \textit{sumabs}=0.52, with 6 non zero row weights and 3 non zero column weights. Then \textit{sumabsu}= 1.56 and \textit{sumabsv}=1.64

\begin{figure}[hbtp]
\centering
\includegraphics[width=9cm]{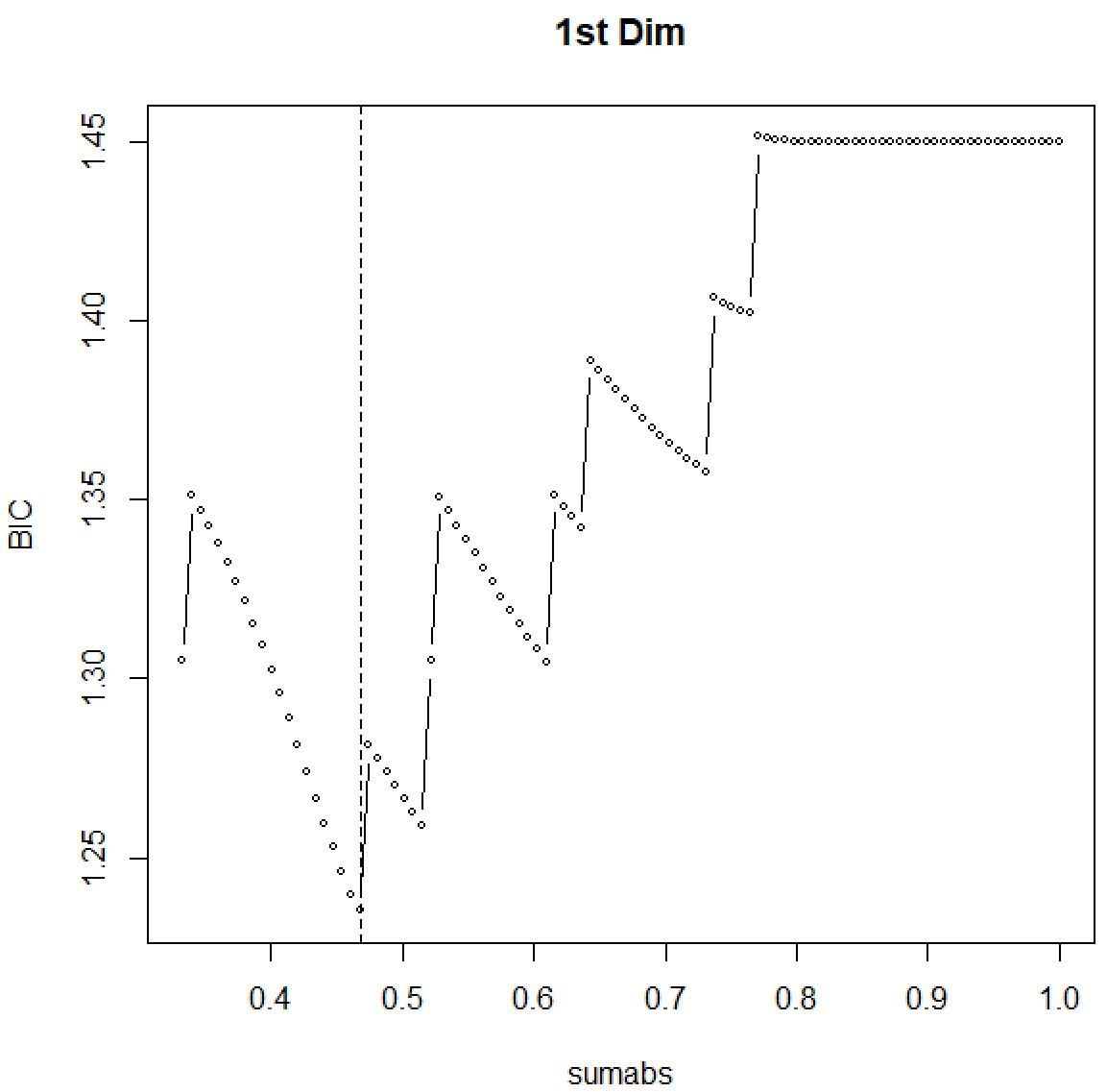}
\caption{BIC as a function of \textit{sumabs}}
\end{figure}

\begin{figure}[hbtp]
\centering
\includegraphics[width=9cm]{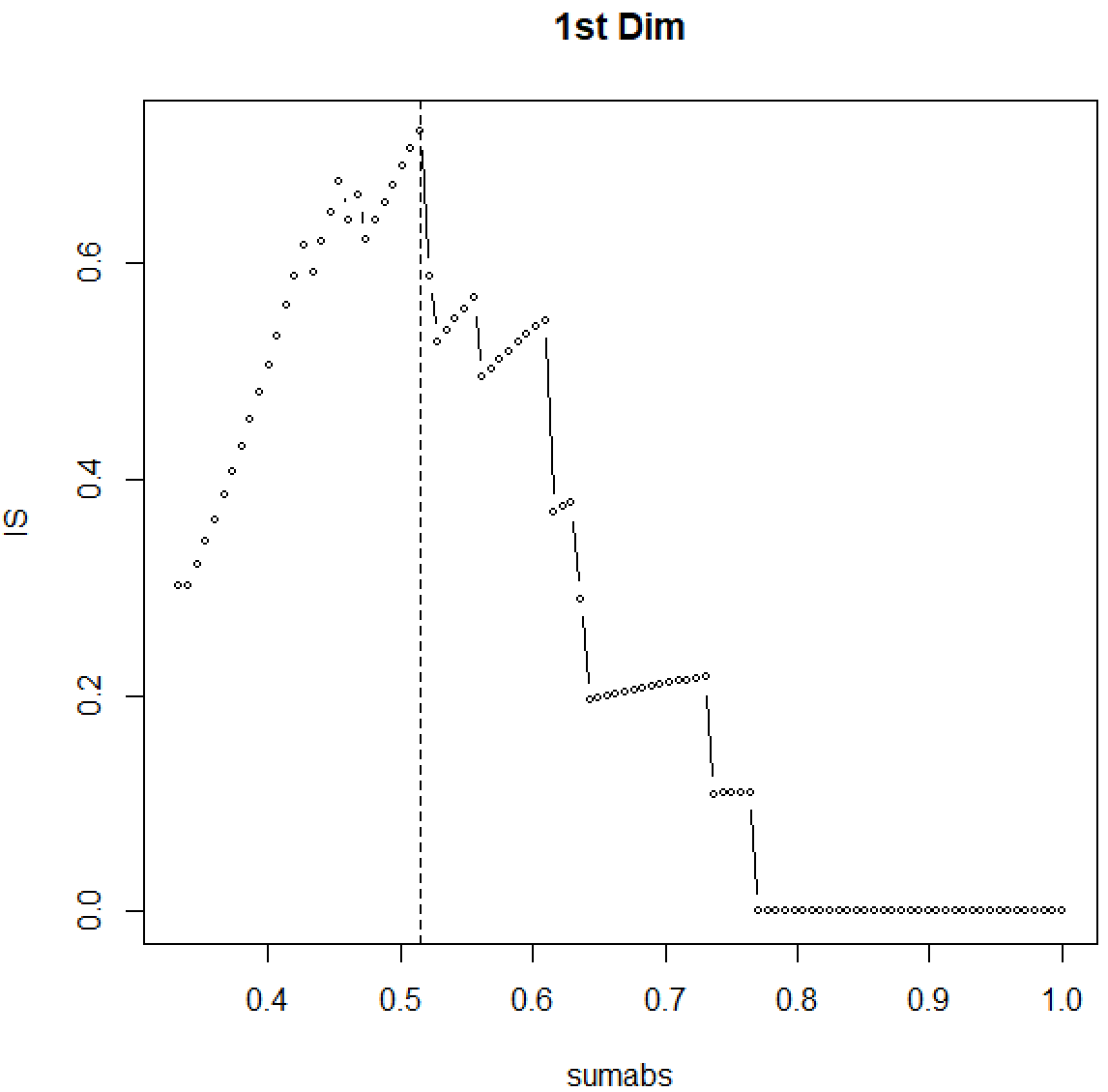}
\caption{IS as a function of \textit{sumabs}}
\end{figure}

Cross validation did not provide satisfactory results: we conducted 10 repeated experiments using \texttt{R} function \texttt{PMD.cv}, the resulting best parameter values were very unstable, as shown in  Table 4. A possible explanation is that it may be caused by the small dimension of the data set. CV will be discontinued for further study.
\newpage 
\begin{center}
Table 4: Unstable “best parameter value” by cross-validation of 10 repeated experiments
\end{center}
\begin{center}
\begin{filecontents*}{Table4.csv}
1,2,3,4,5,6,7,8,9,10
0.58,0.70,0.62,0.33,0.56,0.52,0.51,0.46,0.75,0.74
\end{filecontents*}
\csvautobooktabular{Table4.csv}
\end{center}
\bigskip

b.	Joint optimization of \textit{sumabsu} and \textit{sumabsv}

Let us now look for the “best” values of  \textit{sumabsu} and \textit{sumabsv} with a bidimensional grid. We cannot draw plots similar as Figure 2 to Figure 4, but we may draw various response surfaces and contour plots of BIC and IS  as functions of \textit{sumabsu} and \textit{sumabsv}.
The contour plot of BIC in Figure 8 gives the following optimal pair for the first dimension: \textit{sumabsu }= 1.22,  \textit{sumabsv}=1.44  with BIC = 1.26.

\begin{figure}[hbtp]
\centering
\includegraphics[width=9cm]{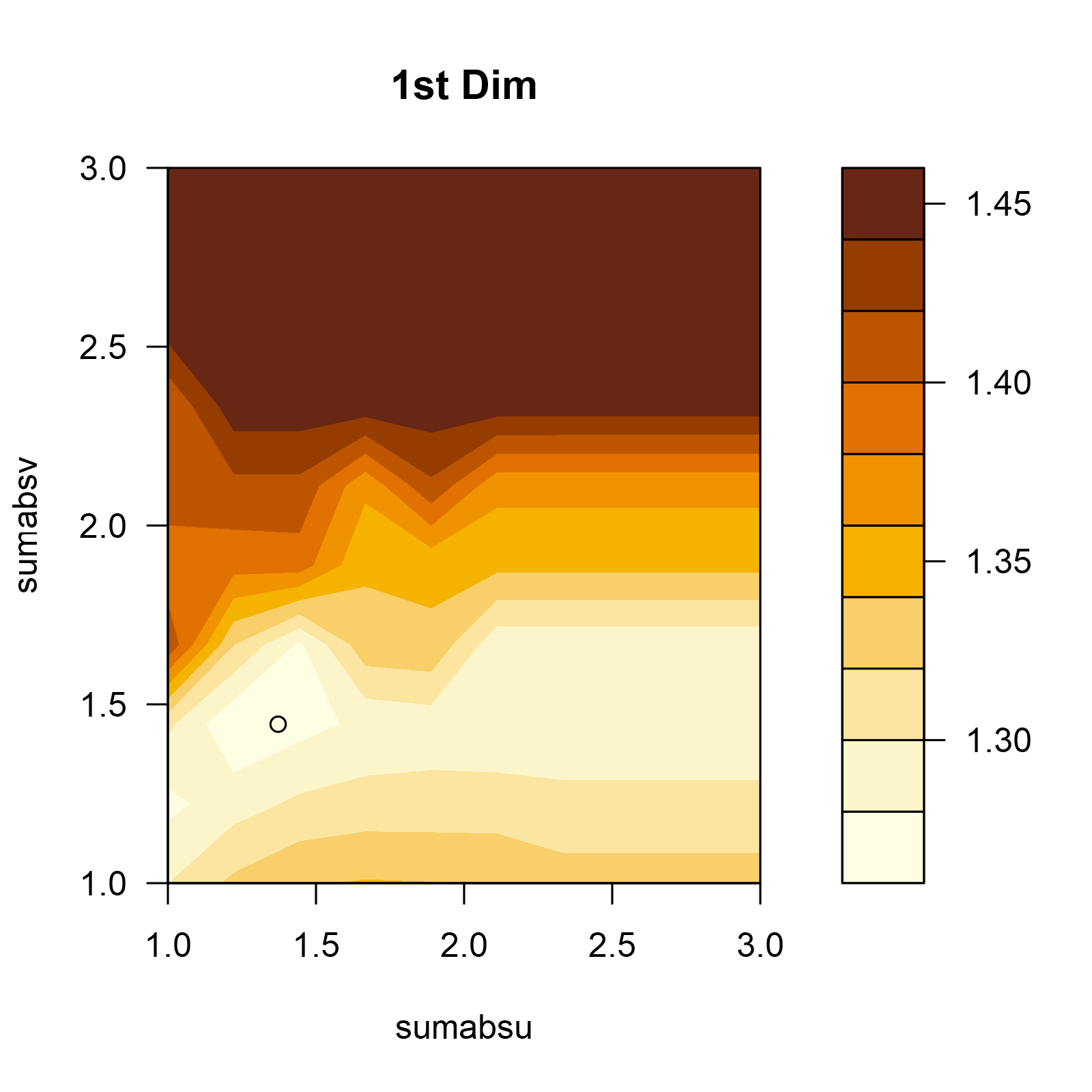}
\caption{Contour plot of BIC as a function of \textit{sumabsu} and \textit{sumabsv} for the first dimension}
\end{figure}

The contour plot of IS (Figure 9) gives the following optimal pair : \textit{sumabsu}= 1.44, \textit{sumabsv}=1.67

\begin{figure}[hbtp]
\centering
\includegraphics[width=9cm]{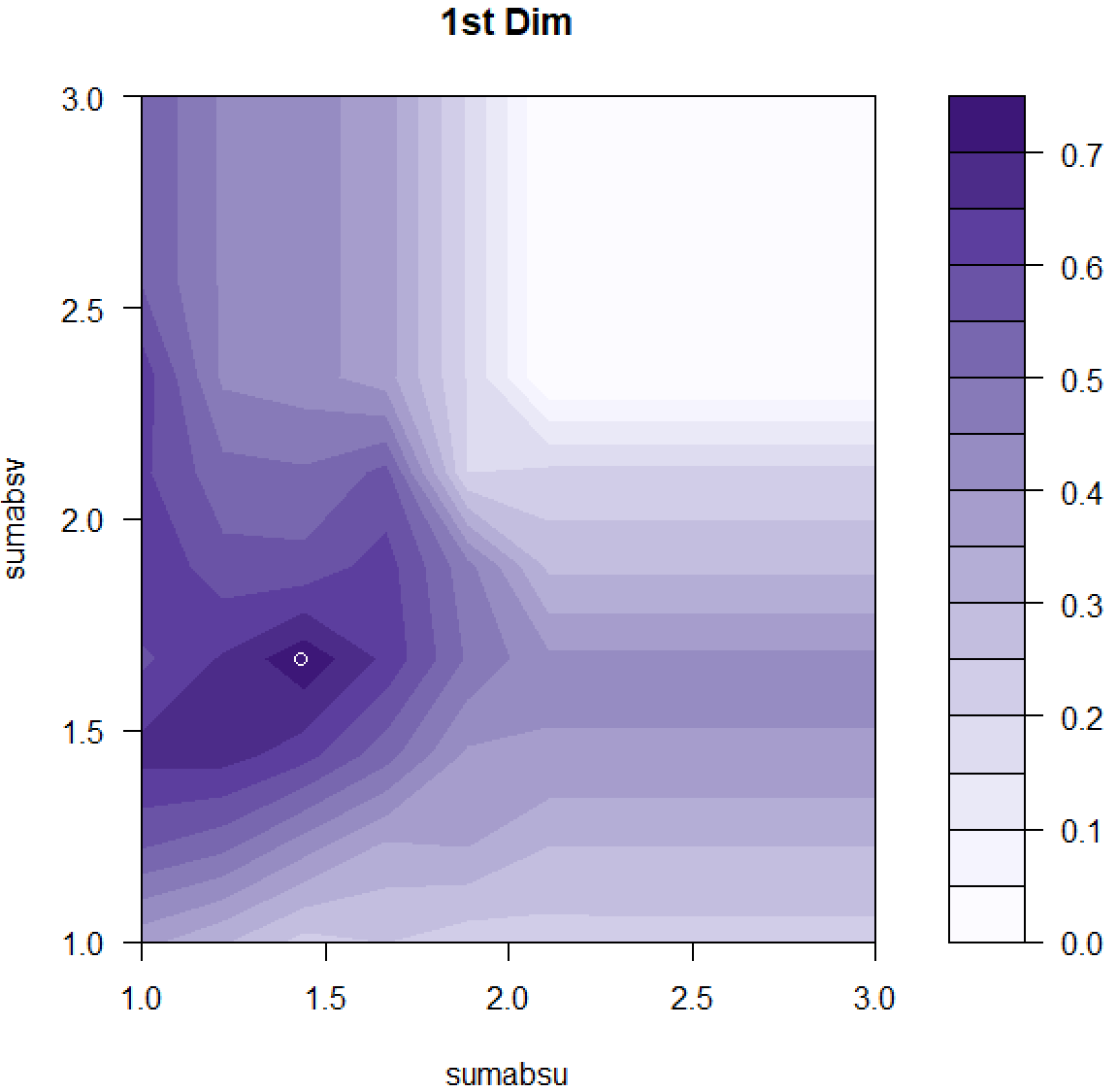}
\caption{Contour plot of IS as a function of \textit{sumabsu} and \textit{sumabsv} for the first dimension}
\end{figure}
\pagebreak 
\subsubsection{Tuning the sparsity parameters for the second dimension}
We will use here our deflation technique pPMD. 

a. BIC fails to give a adequate solution for both ways of selecting the tuning parameters. Given \textit{sumabs}=0.47 for the first dimension, the optimum value for \textit{sumabs} for the second dimension is equal to 0.33 which gives only 2 nonzero weights for the rows and one for the columns (see Figure 10). The solution is too sparse.

\begin{figure}[hbtp]
\centering
\includegraphics[width=11cm]{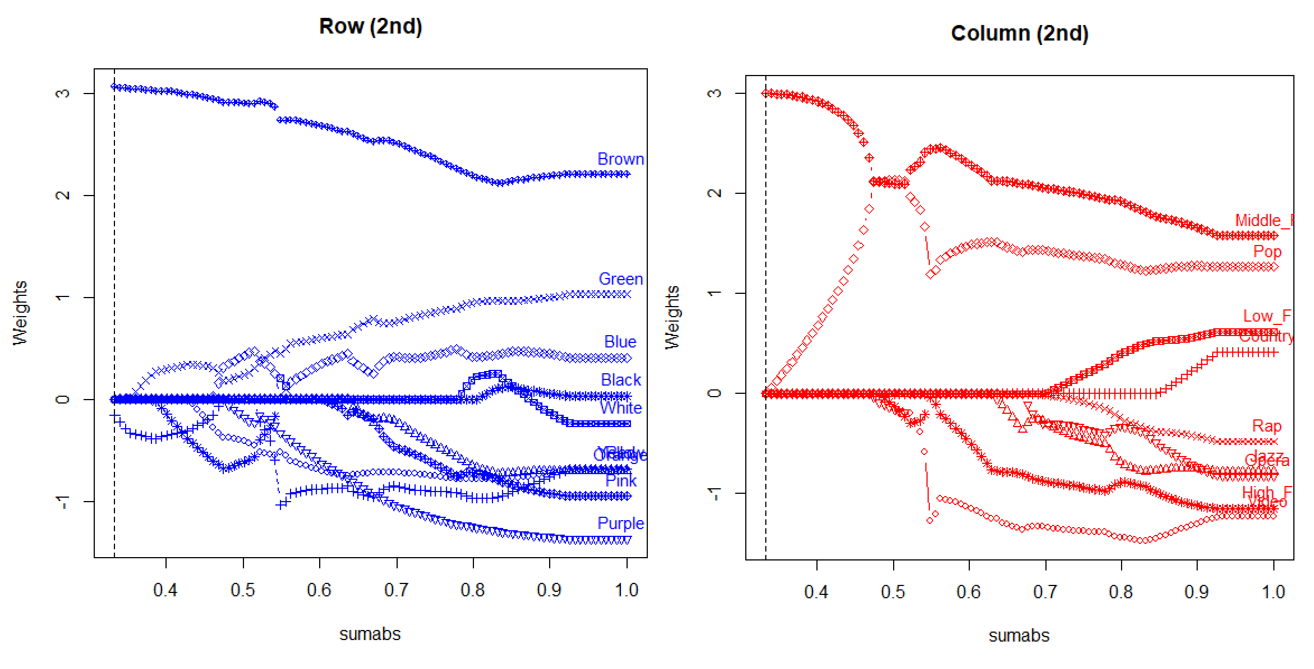}
\caption{Second dimension weights paths when \textit{sumabs1}= 0.47}
\end{figure}

The contour plot for BIC does not give also an acceptable solution for the second sparse dimension, since the optimal values for \textit{sumabsu} and \textit{sumabsv} are equal to 1, as illustrated by Figure 11\pagebreak 

\begin{figure}[hbtp]
\centering
\includegraphics[width=9cm]{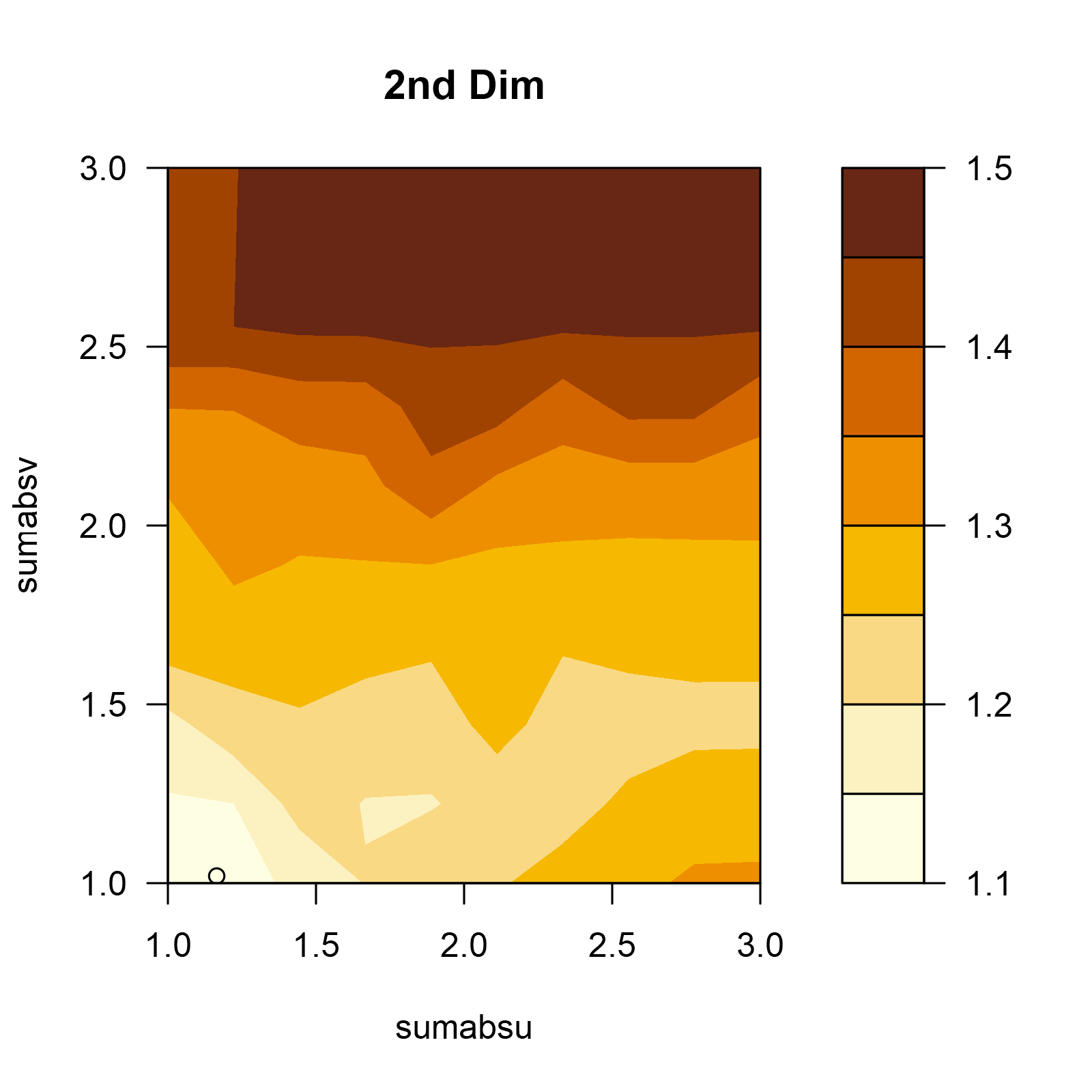}
\caption{contour plot of BIC as a function of \textit{sumabsu} and \textit{sumabsv} for the second dimension}
\end{figure}

b. The IS criterium gives a more satisfactory solution, see Figure 12. The optimal value of \textit{sumabs} for the second dimension is 0.61 which gives respectively 6 non zero weights for the rows and 4 for the columns (Figure 13)  with \textit{sumabsu}=1.83 and \textit{sumabsv}=1.93.

\begin{figure}[hbtp]
\centering
\includegraphics[width=9cm]{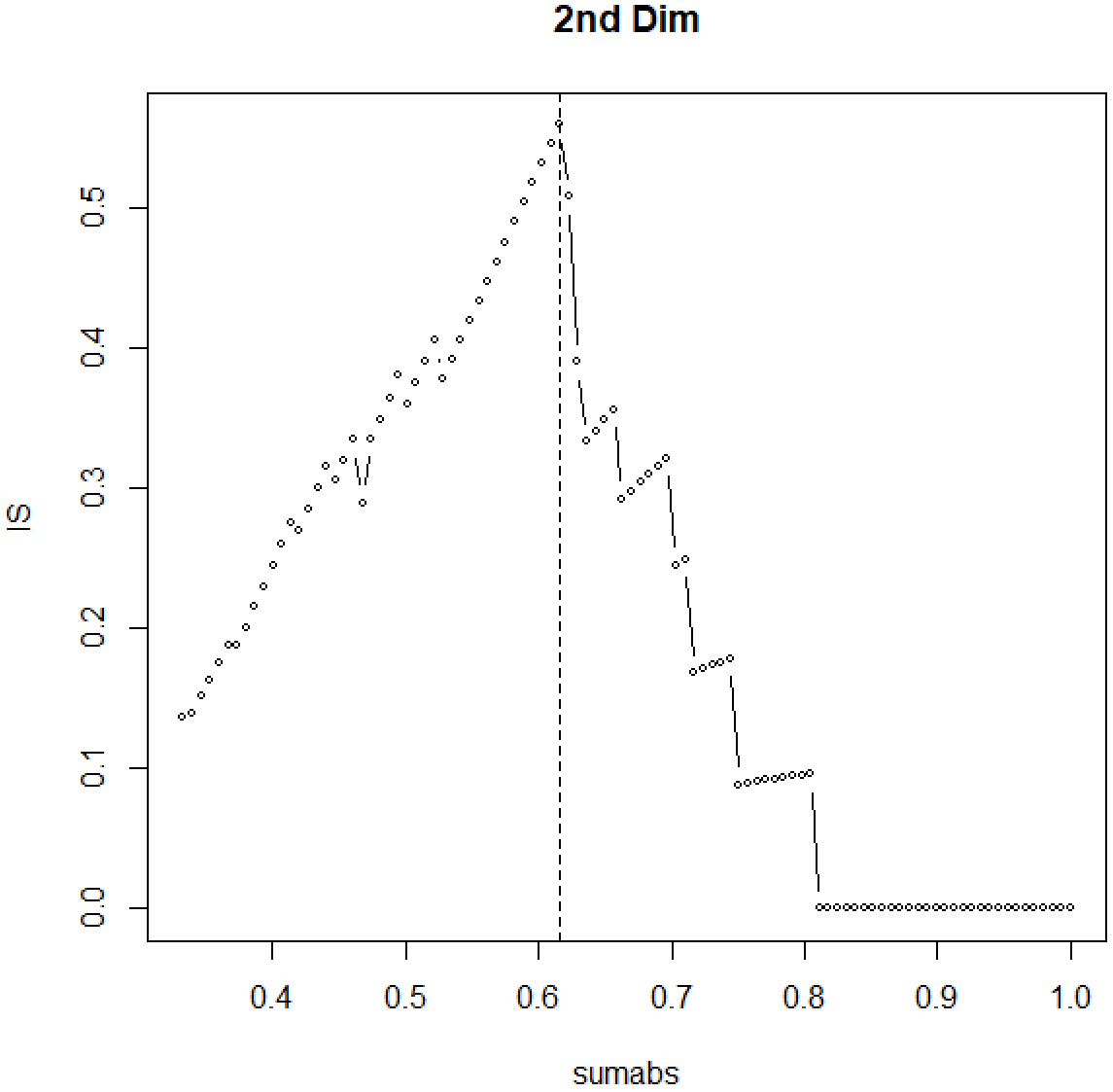}
\caption{Variations of IS for the second dimension}
\end{figure}

\begin{figure}[hbtp]
\centering
\includegraphics[width=9cm]{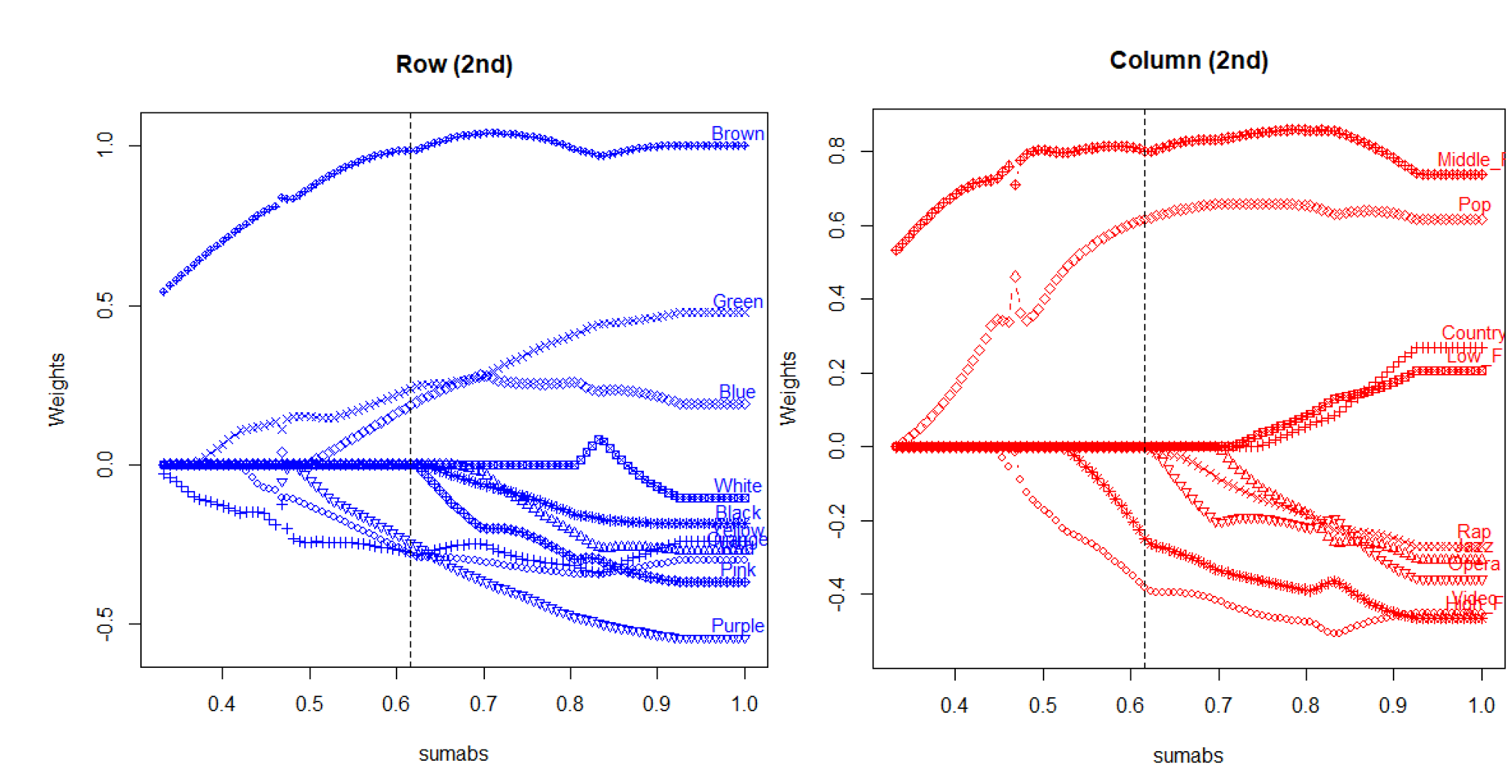}
\caption{Second dimension weights paths when \textit{sumabs1}= 0.52. Dashed vertical line at optimal IS}
\end{figure}

The bidimensional optimization of IS gives the optimal values \textit{sumabsu2}= 1.67, \textit{sumabsv2}=2.11, see Figure 14.

\begin{figure}[hbtp]
\centering
\includegraphics[width=9cm]{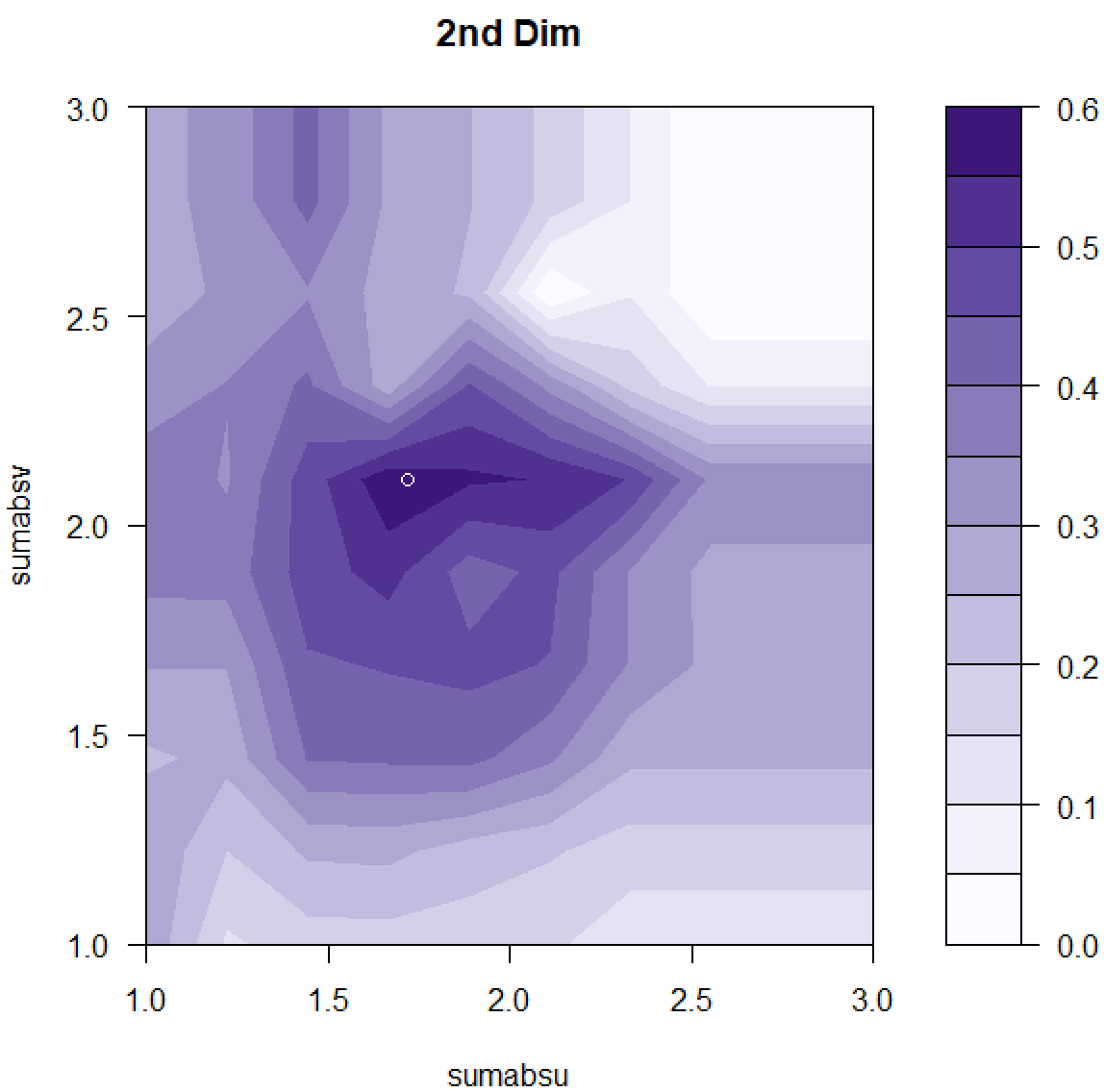}
\caption{Contour plots of IS for the second dimension}
\end{figure}

\pagebreak 
\subsubsection{Sparse CA}
With the previous values of \textit{sumabsu} and \textit{sumabv} for the first two dimensions, one finds a first pseudo eigenvalue ${\lambda _1} = 0.2277$ which accounts for 30.5\% of the variance, and a second pseudo eigenvalue ${\lambda _2} = 0.1175$ which accounts for 18.28\%.

We may note that : 

a.	the percentage of explained variance are a little smaller than in standard CA

b.	the graphical displays are very similar (see Figure 15). 

c.	low contributions have been set to zero, while high contributions are highlighted (see Table 5)

d.	some categories (Red, Green, Blue, Jazz) have zero weights for both dimensions and do not contribute to the display (see Table 5) . They might be deleted but have been kept in the map.

e.	the weight vectors are nearly orthogonal : $ < {{\bf{u}}_1},{{\bf{u}}_2} >  = 0.0085$ and $ < {{\bf{v}}_1},{{\bf{v}}_2} >  =  - 0.0047$

f.	the coordinates vectors are also nearly orthogonal: $ < {{\bf{a}}_1},{{\bf{a}}_2} >  = 0.0128$  and  $ < {{\bf{b}}_1},{{\bf{b}}_2} >  = 0.0320$

\pagebreak 
\begin{center}
Table 5: Weights, contributions and coordinates with separate optimization of \textit{sumabsu} and \textit{sumabsv} by IS criterium
\end{center}
\begin{center}
\begin{filecontents*}{Table5.csv}
,u1,u2,ctr1,ctr2,a1,a2
Red,0,0,0,0,-0.020,0.117
Orange,0.047 ,0,1,0,0.238,0.065
Yellow,0.161 ,0,14,0,0.296,-0.118
Green,0,0,0,0,0.120,-0.251
Blue,0,0,0,0,0.104,-0.231
Purple,0.343 ,0.201,34,19,0.504,0.430
White,0,0.832,0,358,0.216,0.746
Black,-1.202,0,929,0,-1.095,0.091
Pink,0.284,0.233,21,24,0.467,0.460
Brown,0,-0.877,0,598,0.030,-0.717
,,,,,,
,v1,v2, ctr1, ctr2,b1,b2
Video,0.295,0,42,0,0.456,0.146
Jazz,0,0,0,0,0.194,0.013
Country,0.122,-0.345,7,97,0.352,-0.350
Rap,-1.054,0,542,0,-0.914,-0.087
Pop,0,-0.466,0,177,0.208,-0.457
Opera,0,0.639,0,333,0.108,0.600
LowF,-0.915,0,409,0,-0.831,-0.197
HighF,0,0.654,0,349,0.277,0.620
MiddleF,0,-0.235,0,45,0.150,-0.289
\end{filecontents*}
\csvautobooktabular{Table5.csv}
\end{center}
\bigskip

\begin{figure}[hbtp]
\centering
\includegraphics[width=9cm]{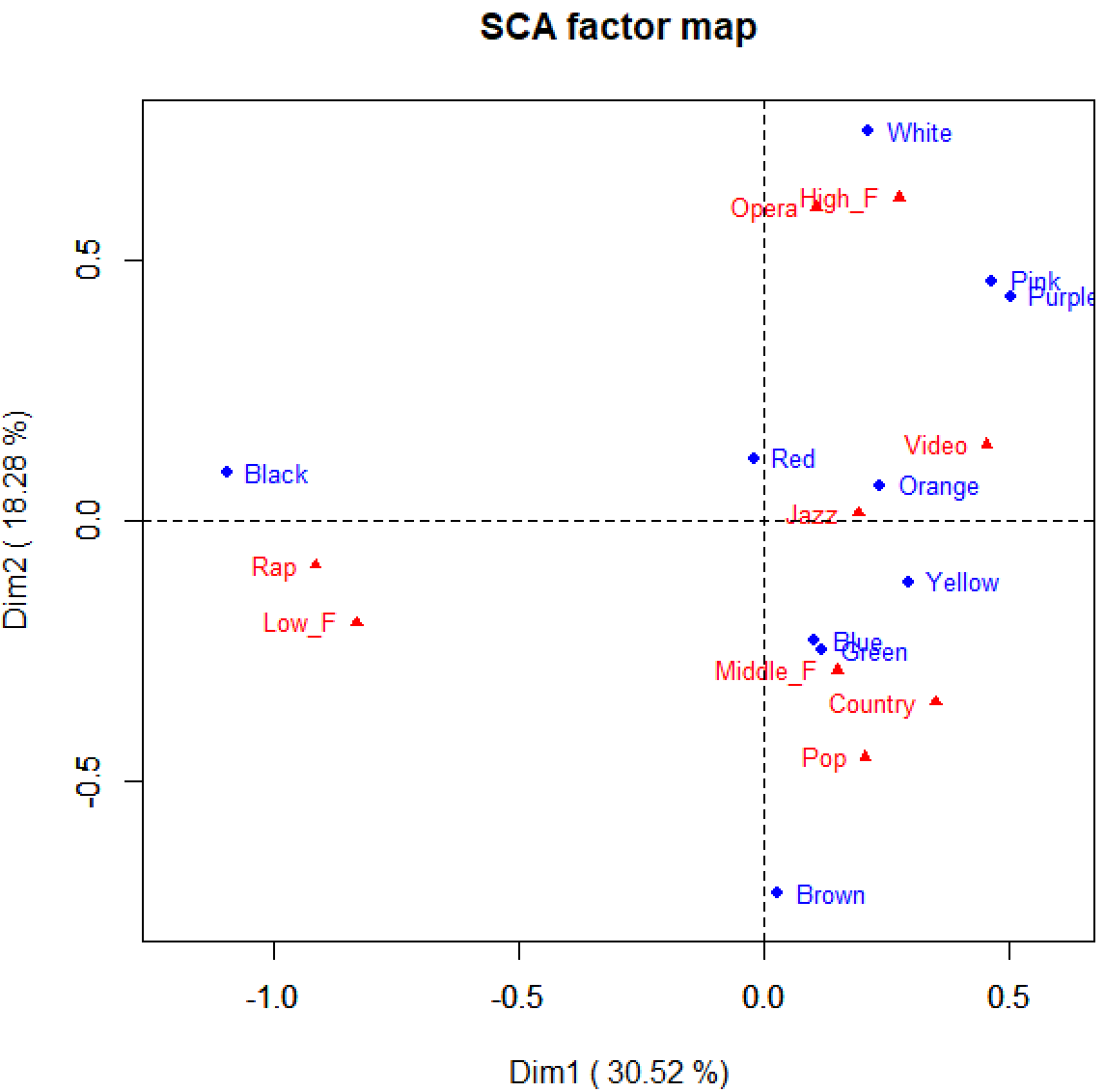}
\caption{Sparse CA with joint optimization of \textit{sumabsu} and \textit{sumabsv}. IS criterium}
\end{figure}
\pagebreak 
The comparison with results where only \textit{sumabs} is optimized shows that the rows weights are less sparse while the columns weights are sparser. The percentages of variance explained by the two dimensions are very close 48.4\% instead of 48.8\%. The graphical display does not show much difference, see Table 6 and Figure 16. 

\pagebreak 
\begin{center}
Table 6: Weights, contributions and coordinates. \textit{sumabs} optimization\\
\medskip 
\begin{filecontents*}{Table6.csv}
,u1,u2,ctr1,ctr2,a1,a2
Red,0,0.281,0,70,-0.044,0.340
Orange,0.114 ,0,5,0,0.236,0.177
Yellow,0.169 ,0.277,16,71,0.261,0.344
Green,0,-0.239,0,49,0.101,-0.313
Blue,0,-0.187,0,25,0.137,-0.294
Purple,0.411 ,0.245,49,29,0.508,0.388
White,0.132 ,0,5,0,0.270,-0.100
Black,-1.166,0,887,0,-1.082,0.119
Pink,0.369 ,0,37,0,0.480,0.229
Brown,0,-0.988,0,757,0.030,-0.809
,,,,,,
,v1,v2, ctr1, ctr2,b1,b2
Video,0.215,0.383,23,119,0.441,0.455
Jazz,0,0,0,0,0.188,0.213
Country,0,0,0,0,0.314,-0.106
Rap,-1.062,0,558,0,-0.908,0.189
Pop,0,-0.617,0,309,0.181,-0.537
Opera,0,0,0,0,0.155,0.234
LowF,-0.920,0,419,0,-0.830,-0.197
HighF,0,0.250,0,51,0.318,0.383
MiddleF,0,-0.800,0,520,0.142,-0.634
\end{filecontents*}
\csvautobooktabular{Table6.csv}
\end{center}
\bigskip
\pagebreak 

\begin{figure}[hbtp]
\centering
\includegraphics[width=9cm]{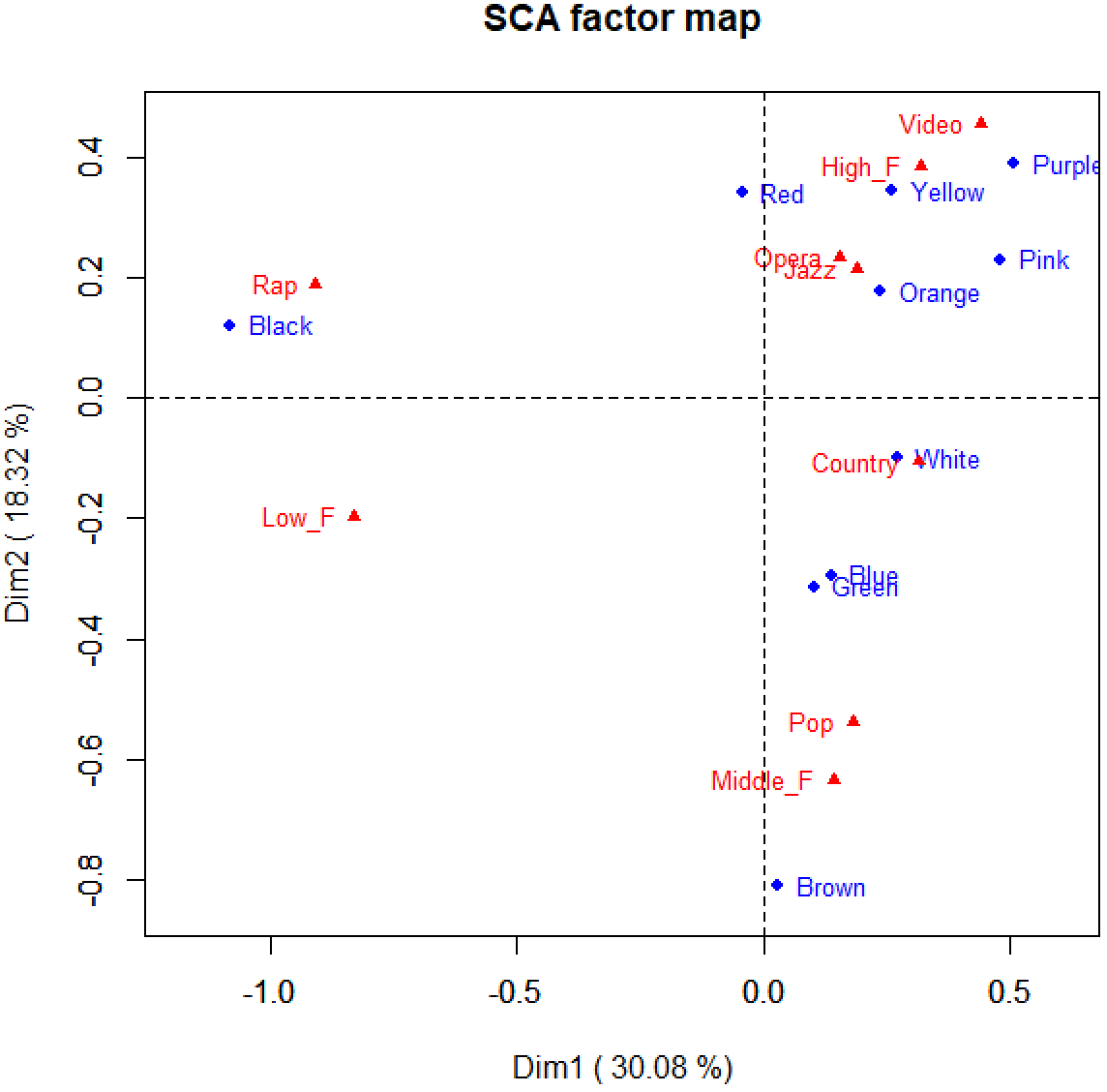}
\caption{Sparse CA; \textit{sumabs} optimization}
\end{figure}
\

\subsubsection{Summary}
What have we learned from this simple example ?

-   Sparse CA facilitates the reading of important categories

-	pPMD provides near orthogonal solutions

-	BIC and cross-validation fail to select the sparsity parameters, while IS performs well 

-	Joint optimization of \textit{sumabsu} and \textit{sumabsv} gives slightly better percentage of explained variance, and very similar displays compared to the optimization of \textit{sumabs}

\section{Application to textual data}
CA plays a specific part in the field of text mining, also called text analytics. As an exploratory tool, it is a multidimensional technique of visualization, not the only one, but one of the most powerful, besides more popular and basic techniques. It must be noted that the first applications of CA by Jean-Paul Benzécri were related to linguistics (\textit{cf.} Lebart, Salem and Berry, 1998). For a recent and comprehensive survey (in french) see  Lebart, Pincemin and Poudat (2019).  Bécue-Bertaut (2019) covers the main multidimensional methods in textual statistics together with the  specially-written \texttt{R} package  \texttt{Xplortext}.

We illustrate sparse CA on the following  data set: the speeches of 43\footnote{There are 45 presidents, but the speech data of presidents William Henry Harrison and James Garfield are missing.}  presidents of the United States (from G.Washington to D.Trump, see Table in appendix). The complete speeches are available at \url{https://www.usa-presidents.info/union/}. 
A related paper (Savoy, 2015) describes a clustering and authorship attribution study over the State of the Union addresses delivered by 41 presidents from 1790 to 2014 .

We first retain the 934 most frequent words that appear more than 220 times in the speeches. Then 207 meaningless words, such as “of”, “the”, etc. are deleted. Finally 572 stems  are extracted from the 727 remaining graphical forms, hence a data table  $ \textbf{N} $ where \textit{I}= 43, \textit{J}= 572.
After a  standard correspondence analysis (CA), a column sparse only correspondence analysis is performed, since we would like to highlight the most significant words, while keeping all presidents. 
\subsection{Standard CA}
There are 42 eigenvalues, most of them being very small. Table 7 and Figure 17 show two dominant eigenvalues, which justifies the use of two-dimensional displays.

\begin{center}
Table 7: The first 10 eigenvalues\\
\medskip 
\begin{filecontents*}{Table7.csv}
Rank,Eigenvalue,Percent,Cum.percent
1,0.196,37.753,37.753
2,0.045,8.713,46.465
3,0.037,7.198,53.664
4,0.025,4.827,58.491
5,0.018,3.530,62.021
6,0.017,3.203,65.224
7,0.014,2.728,67.952
8,0.011,2.101,70.052
9,0.010,2.015,72.067
10,0.009,1.790,73.858
\end{filecontents*}
\csvautobooktabular{Table7.csv}
\end{center}
\bigskip

\begin{figure}[hbtp]
\centering
\includegraphics[width=9cm]{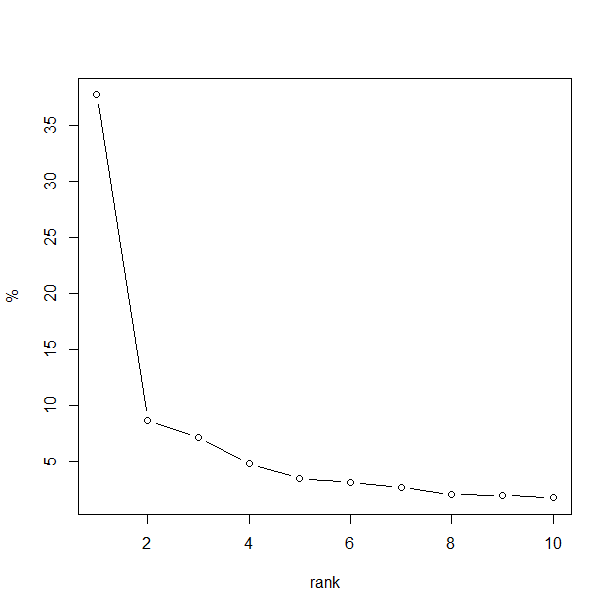}
\caption{Scree plot of the first 10 eigenvalues}
\end{figure}

Due to the large number of words the factor map (Figure 18) is difficult to interpret.
 
\begin{figure}[hbtp]
\centering
\includegraphics[width=9cm]{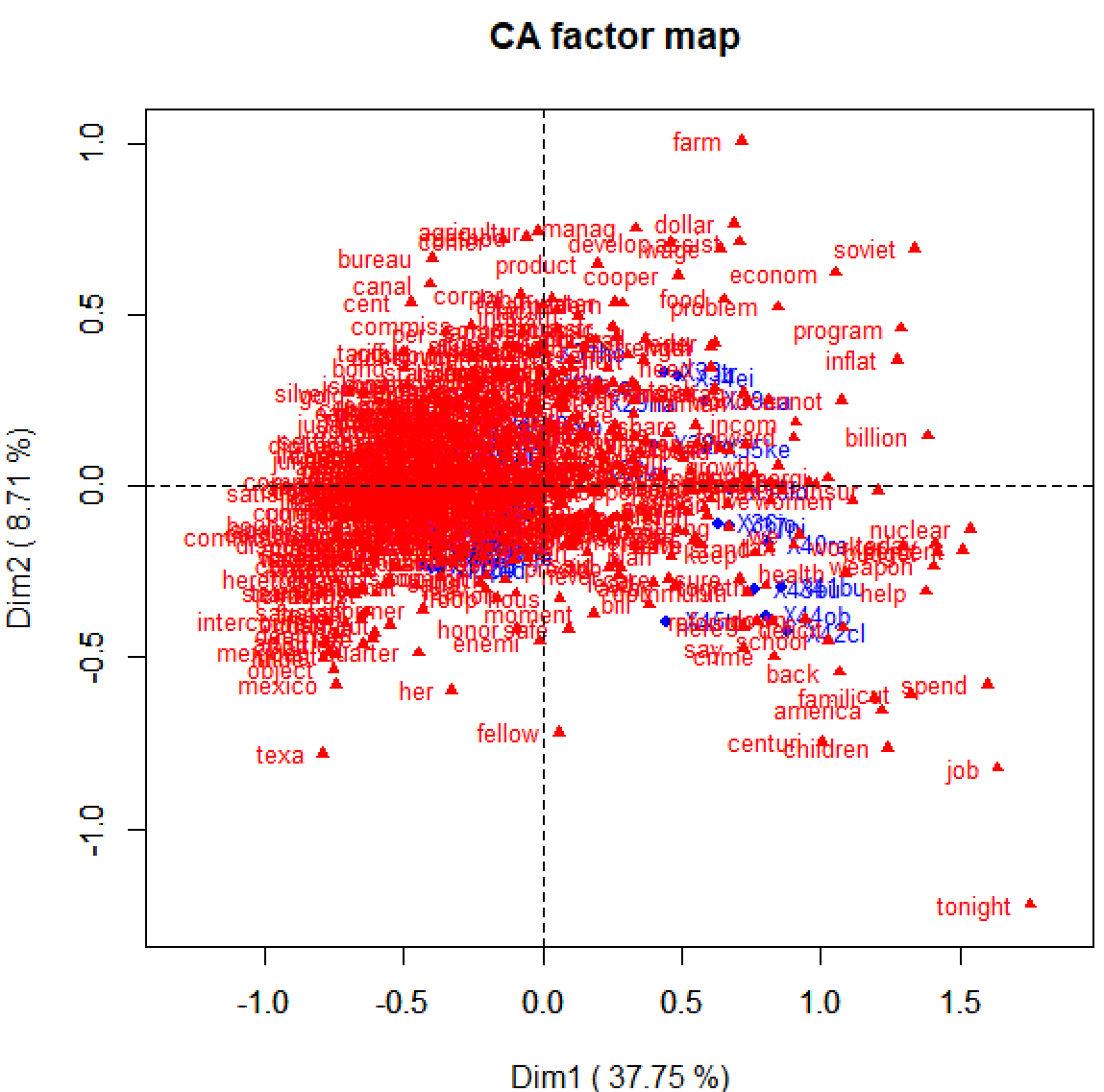}
\caption{CA symmetric map}
\end{figure}

A cluster analysis (Ward's algorithm) of the presidents does not reveal a clear structure : 5 or 8 clusters (Figure 19)

\begin{figure}[hbtp]
\centering
\includegraphics[width=9cm]{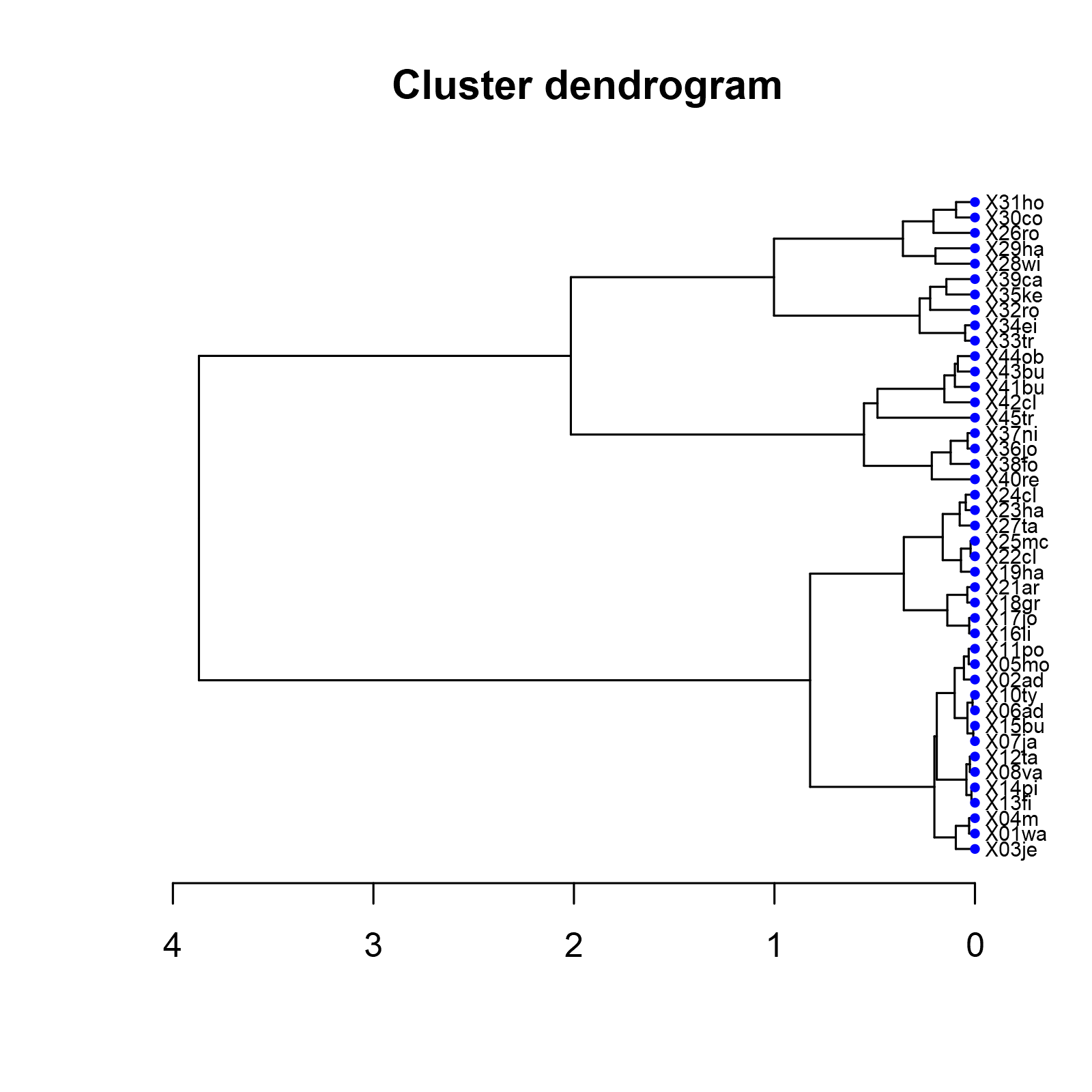}
\caption{Hierarchical clustering of the 43 presidents on the first two CA coordinates }
\end{figure}
\pagebreak 
\subsection{Sparse CA}
We look now for a sparse CA where only the columns (words) are sparsified. It implies that there is no sparsity constraint on $sumabsu = \sum\limits_{i = 1}^I {\left| {{u_i}} \right|} $ and that $sumabsv = \sum\limits_{j = 1}^J {\left| {{v_j}} \right|} $ is the only parameter to choose for each dimension. However optimizing IS criterium may lead to a too large number of non zero coefficients. A pragmatic solution, where the number of non zero coefficients is defined beforehand, may be preferred.

\subsubsection{First dimension}
Figure 20 gives the plot of the IS criterium and of the number of non zero weights against \textit{sumabsv},

With a grid search between 1 and $\sqrt J\simeq 24 $ by 0.2, the optimal value of \textit{sumabsv} which gives the maximal IS value is 11.4. This value corresponds to 288 non zeros which is not sparse enough. So we choose \textit{sumabsv} = 4.2 to have a  number of non zeros equal to 50.

Below is the list of the 50 words with non zero weights, sorted by their contributions. 
 
Positive coordinates (40 words):
  we,     america,  program,  tonight,  help,   our,    must,    world,  budget,  I ,      
  job,  children,  famili,  american,  tax,  today,  billion,  health, percent, econom,  
  spend, new, school,  freedom,  work,   live,  economi,  need, nuclear,  cut,     
  million, togeth,  will,  weapon,   worker,   cannot,   back,     problem,  feder,    down
  
Negative coordinates (10 words): 
  state, unit,  subject, duti,  govern, general,  treati,  public, present, territori

\begin{figure}[hbtp]
\centering
\includegraphics[width=11cm]{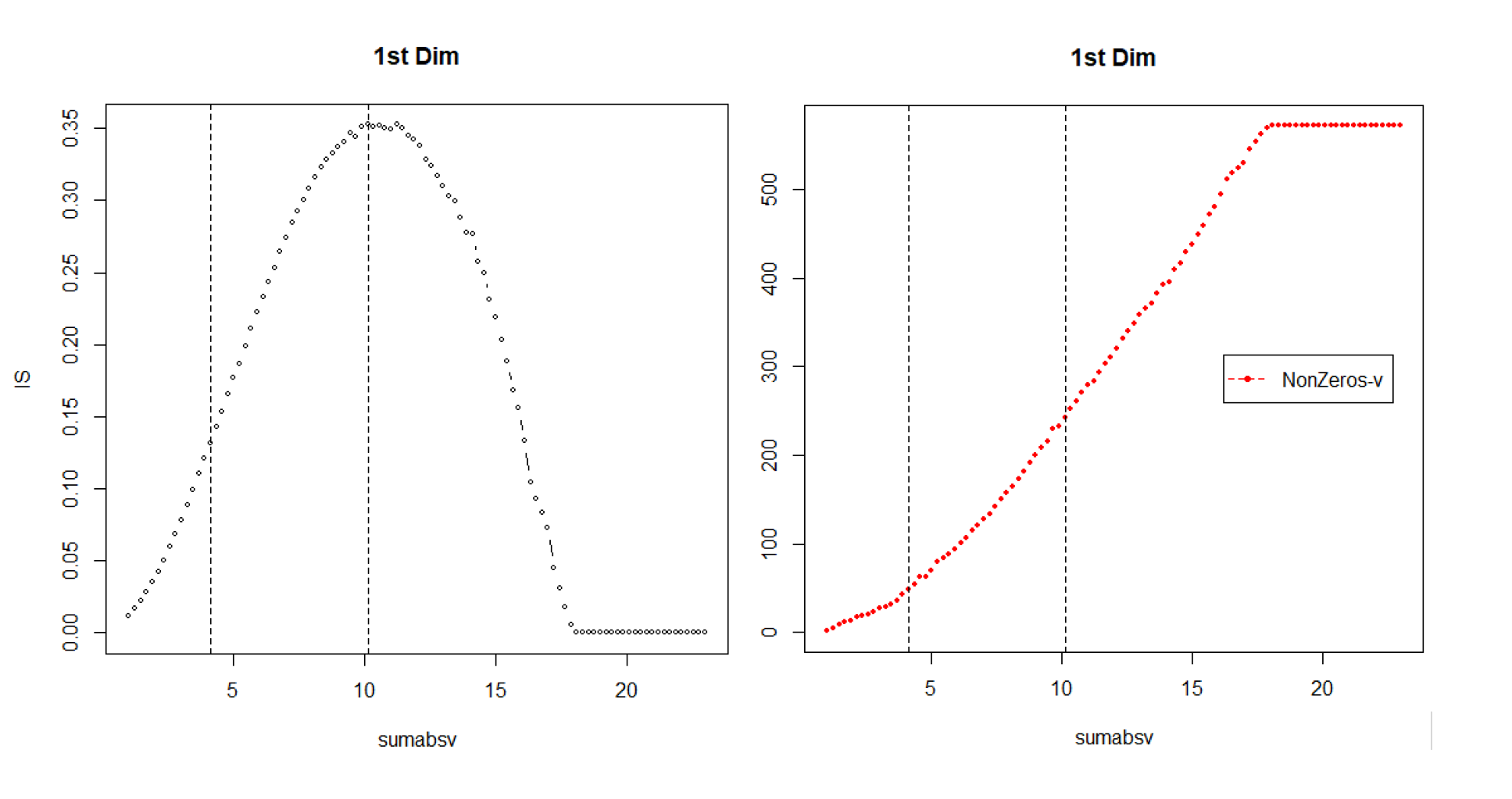}
\caption{IS and the number of non zero weights against \textit{sumabsv} for Dim1 }
\end{figure}

\subsubsection{Second dimension}
Deflating with pPMD  we obtain Figure 21. The optimal value of \textit{sumabsv} is 11 which does not give a sparse enough solution with 279 non zero coefficients.  We decide to choose \textit{sumabsv}=5.2 which corresponds to 51 non zero coefficients.

Below is the list of the 51 words with non zero weights, sorted by their contributions. 37 are common with the first dimension.

Positive coordinates (38 words):
program,      econom,       soviet,       dollar,       feder,        world,        farm,    administr,    develop,      nation,     defens,       price,        assist,       strength,     problem,      polici,       inflat,       major,        product,      manag,       cooper,       effort,       agricultur,   billion,      free,         human,        fiscal,       need,         food,      resourc,      energi,       continu,      militari,     essenti,      legisl,       intern,       peac,    expenditur 

Negative coordinates (13 words): 
tonight,    america,    children,   I,    centuri,    american,   school,    her,    care,  crime,   famili,     say,    from   

\begin{figure}[hbtp]
\centering
\includegraphics[width=11cm]{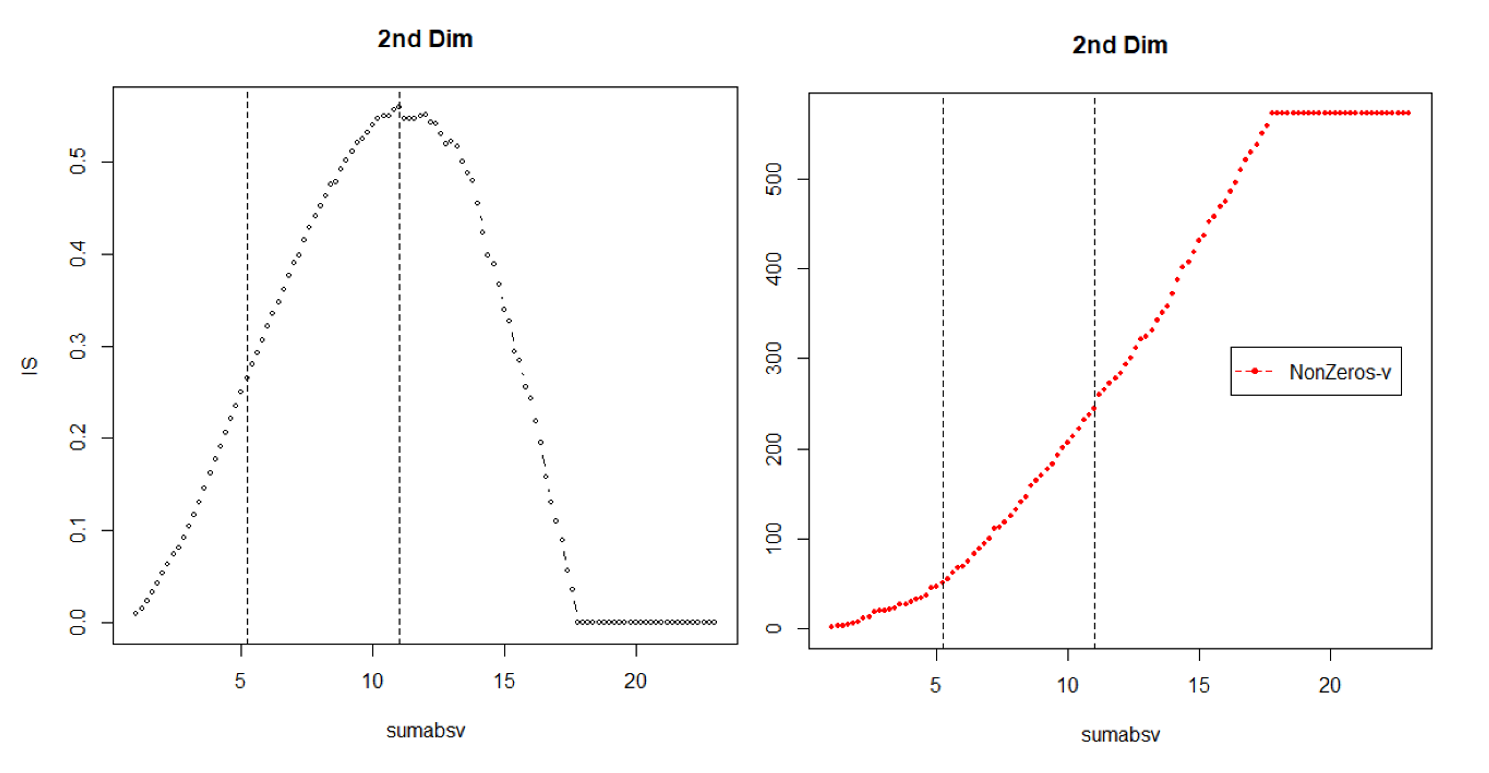}
\caption{IS and the number of non zero weights against \textit{sumabsv} for Dim2 }
\end{figure}

Figure 22 shows the two-dimensional simultaneous plot of presidents and words, where only the words with non zero contributions for both dimensions are displayed. The percentages of variance accounted for both dimensions decrease dramatically compared to standard CA but the interpretation is easier and there is a good correlation between CA and SCA presidents coordinates: $r=0.984$ for the first dimension.

\begin{figure}[hbtp]
\centering
\includegraphics[width=9cm]{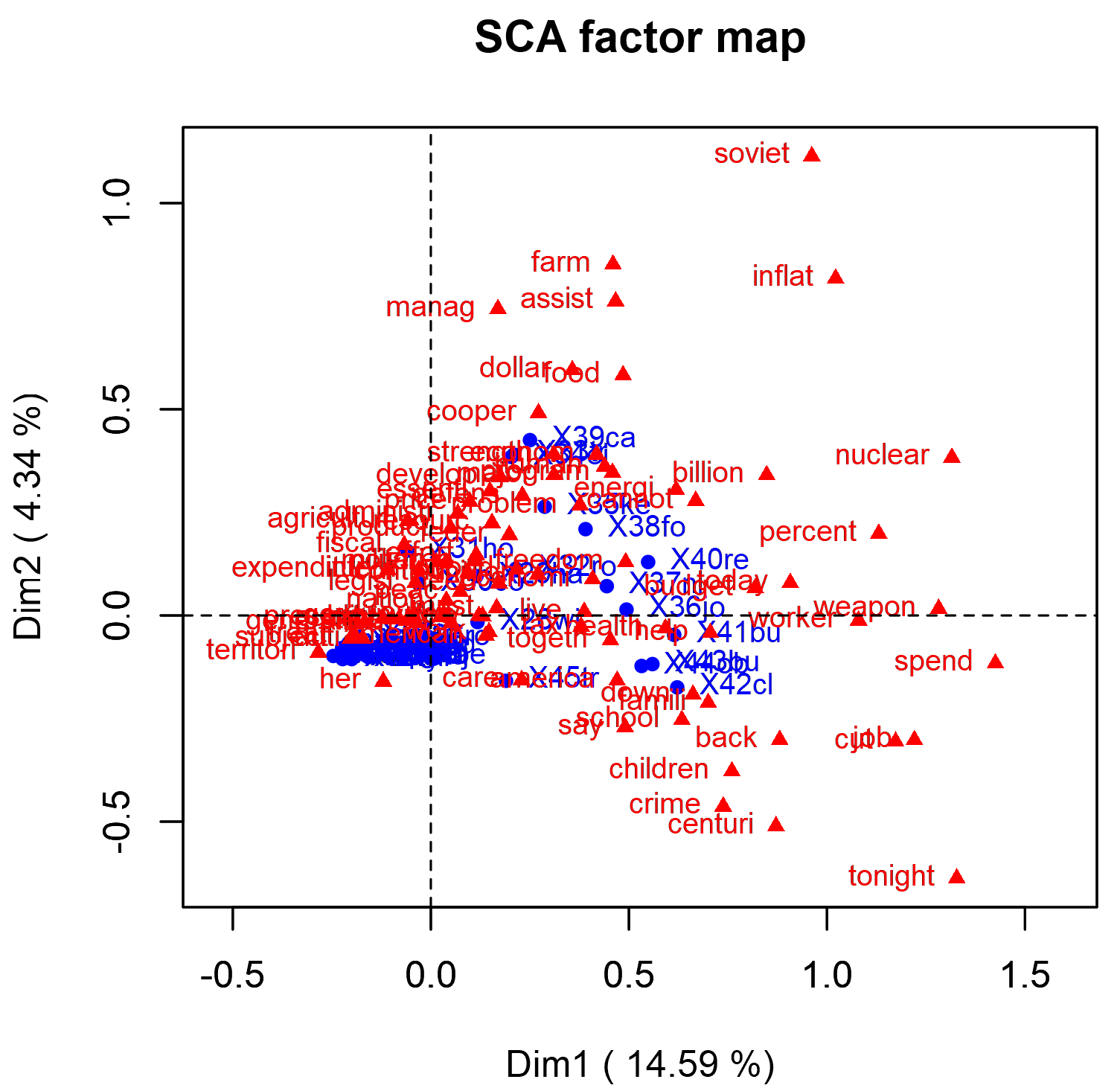}
\caption{SCA factor map: a \& b, without zero-contribution items}
\end{figure}

\pagebreak 

\subsubsection{Cluster analysis}

The clustering of rows (Figure 23) is also clearer. Eight clusters appear where one can see that president Trump (X45tr) looks very different from the other presidents and forms a  class by himself. 

\begin{figure}[hbtp]
\centering
\includegraphics[width=9cm]{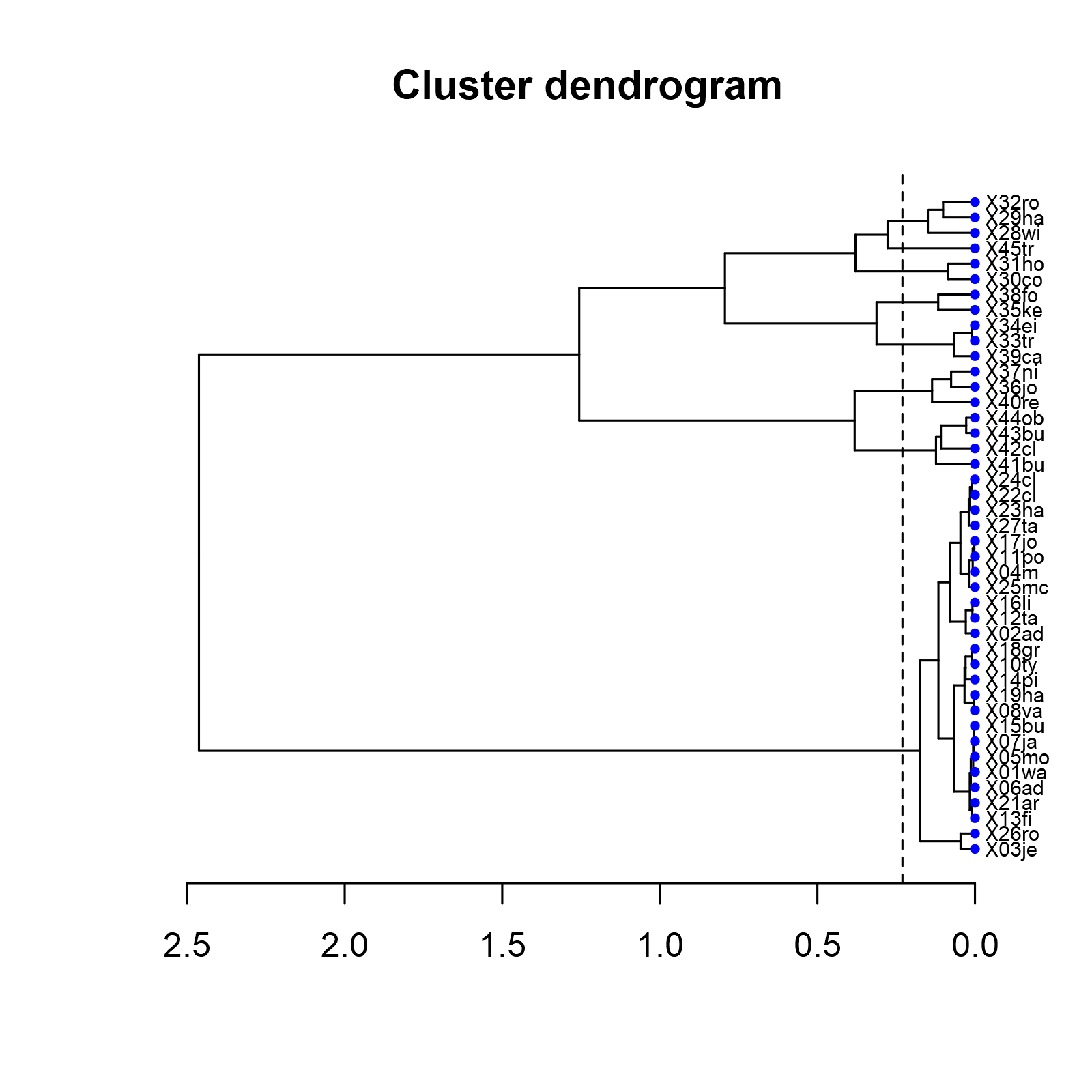}
\caption{Hierarchical clustering of the 43 presidents on the first two sparse CA coordinates }
\end{figure}


Figure 24 provides a display of the 8 clusters projected onto the first two sparse dimensions; each cluster is illustrated with its 3 most typical words. 

The typical words are determined as follows: let \textbf{K} be the contingency table crossing clusters and words. \textbf{K} is here a table with 8 rows and 87 columns, with total \textit{k}. Word \textit{j} is over-represented in cluster \textit{i} (and thus typical) if its conditional relative frequency  $\frac{{{k_{ij}}}}{{{k_i}}}$ is much larger than its marginal frequency $\frac{{{k_j}}}{k}$. If these two frequencies were equal, $k_{ij}$  could be compared to a binomial distribution $B\left( {{k_i},\frac{{{k_j}}}{k}} \right)$.   We keep for each cluster \textit{i} the 3 words with the highest standardized values  $z_{ij}$:

\[{z_{ij}} = \frac{{\left( {{k_{ij}} - \frac{{{k_i}{k_j}}}{k}} \right)}}{{\sqrt {\frac{{{k_i}{k_j}}}{k}\left( {1 - \frac{{{k_j}}}{k}} \right)} }}\]

\begin{figure}[hbtp]
\centering
\includegraphics[width=9cm]{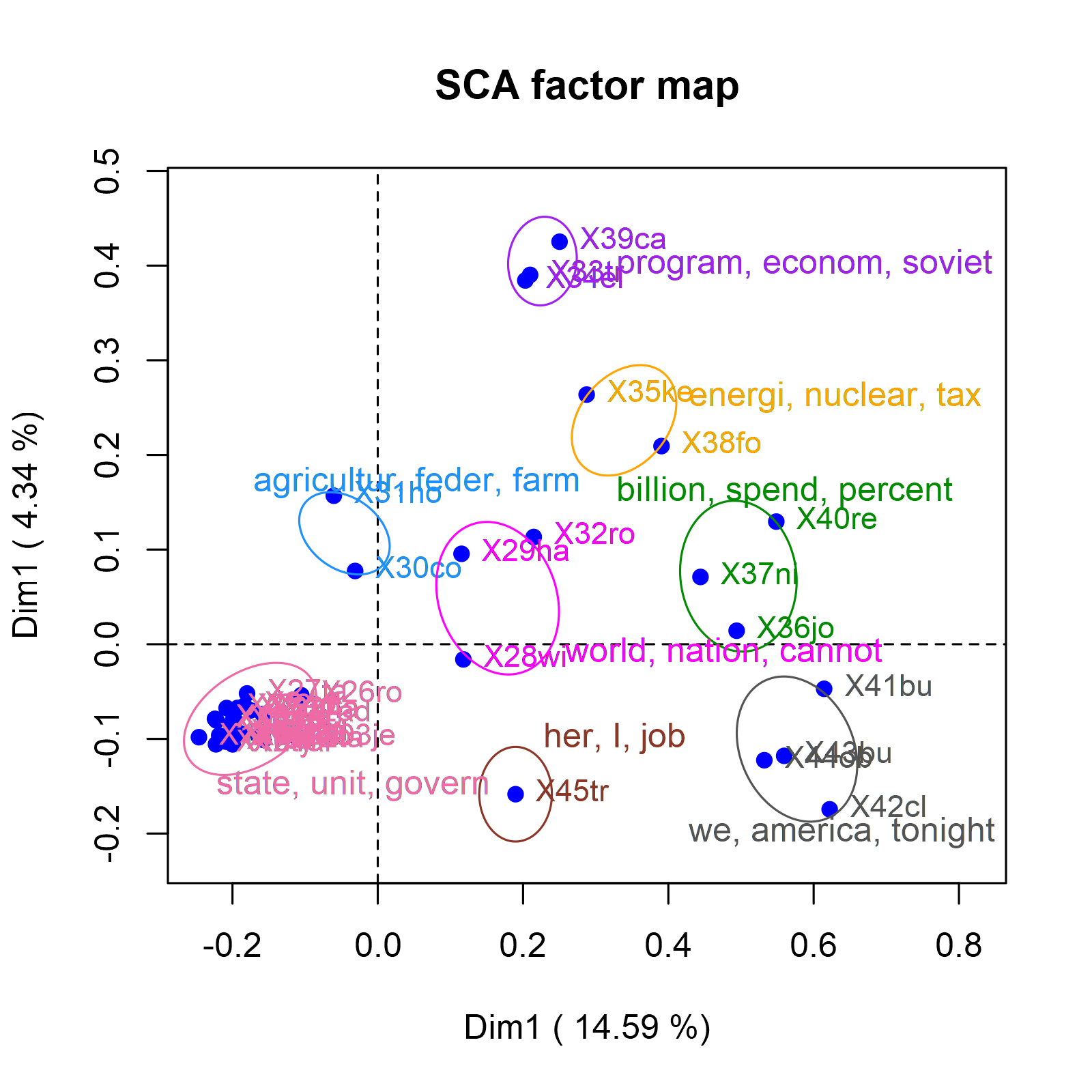}
\caption{Clusters and words on the first two sparse CA coordinates }
\end{figure}
\pagebreak

\section{Conclusion and perspectives}
We have proposed a sparse version of correspondence analysis that highlights the most important categories that determine the underlying axes of dependence between two categorical variables. We believe that this method is particularly well suited to the case where at least one (if not both) variables has a high number of categories as we have illustrated on textual data. The method is flexible and allows different levels of sparsity for rows and columns. A simple and efficient deflation technique pPMD has been proposed.

As with other sparse methods, however, the search for solutions is carried out at the cost of losing properties characteristic of correspondence analysis: orthogonality, barycentric relations. The use of components (or coordinates) rather than weights, however, allows for graphical representations that can be interpreted in a similar way to that of standard correspondence analysis.
The search for an optimal level of sparsity remains a delicate problem and for large contingency tables, the pre-specification of the number of zeros is recommended.

Among the perspectives raised by this new method, two seem particularly interesting to us:

-	since pairs of sparse factors determine sub-matrices of $\textbf{N} $,  are there some links with biclustering, also known as “coclustering” or “block clustering”, \textit{cf} Govaert  \& Nadif (2013) where one looks for clustering simultaneously rows and columns?

-	 even if only one of the two sets (rows or columns) is sparsified,  sparse correspondence analysis remains basically a symmetrical method where rows and columns play the same role. In future work, we plan to develop a sparse variant of the non symmetric correspondence analysis introduced by Lauro \& D’Ambra (1984). \textit{Cf.} also D’Ambra \& Lauro (1992).

\section*{References}

\qquad     Abdi H., Béra M. (2017) Correspondence Analysis. In: Alhajj R., Rokne J. (eds) \textit{Encyclopedia of Social Network Analysis and Mining.} Springer, New York, NY 

Adachi, K., Trendafilov, N.T. (2015) : Sparse principal component analysis subject to prespecified cardinality of loadings. \textit{Computational Statistics} \textbf{31}, 1–25

D’Ambra, L., Lauro, N. C. (1992). Non symmetrical exploratory data analysis.\textit{ Statistica Applicata}, \textbf{4}(4), 511-529.

Bécue-Bertaut, M. (2019): \textit{Textual Data Science with }\texttt{R}, Chapman and Hall/CRC

Beh, E. J., Lombardo, R. (2014). \textit{Correspondence analysis: theory, practice and new strategies}. John Wiley \& Sons.

Bernard, A., Guinot, C., Saporta, G. (2012): Sparse principal component analysis for multiblock data and its extension to sparse multiple correspondence analysis. In: Colubi, A., Fokianos, K., Gonzalez-Rodriguez, G., Kontoghiorghes, E. (eds.) \textit{Proceedings of 20th International Conference on Computational Statistics (COMPSTAT 2012)},  99–106 

Fisher R.A. (1940): The precision of discriminant functions. \textit{Ann. Eugen}., \textbf{10}, 422-429.

Govaert, G., Nadif, M. (2013). \textit{Co-clustering: models, algorithms and applications}. John Wiley \& Sons.

Greenacre, M. J. (2010). Correspondence analysis. \textit{Wiley Interdisciplinary Reviews: Computational Statistics}, \textbf{2}(5), 613–619. \url{doi:10.1002/wics.114}

Guillemot V, Beaton D, Gloaguen A, Löfstedt T, Levine B, Raymond N, \textit{et al}. (2019) A constrained singular value decomposition method that integrates sparsity and orthogonality. \textit{PLoS ONE} \textbf{14}(3): e0211463. \url{https://doi.org/10.1371/journal.pone.0211463}

Hall, P., Marron, J. S., Neeman, A. (2005) Geometric representation of high dimension, low sample size data. \textit{J. R. Stat. Soc. Ser. B Stat. Methodol}. \textbf{67},   3, 427–444.

Jolliffe, I.T., Trendafilov, N.T. and Uddin, M. (2003) A modified principal component technique based on the LASSO. \textit{Journal of Computational and Graphical Statistics}, \textbf{12}, 531–547 

Laclau C., Nadif, M. (2015) Diagonal Co-clustering Algorithm for Document Word Partitioning. In: Fromont E., De Bie T., van Leeuwen M. (eds) \textit{Advances in Intelligent Data Analysis XIV. IDA 2015. Lecture Notes in Computer Science},  \textbf{9385}, 170-180, Springer, Cham

Lauro, N.,  D’Ambra, L. (1984). L’analyse non symétrique des correspondances. \textit{Data analysis and informatics III}, (E.Diday \textit{et al}. editors), 433-446. North-Holland.

Lebart, L., Salem, A., Berry, L. (1998) \textit{Exploring textual data}, Kluwer

Lebart, L., Pincemin, B., Poudat, C. (2019) \textit{Analyse des données textuelles}, Presses de l’Université du Québec

Lebart, L., Saporta, G. (2014) Historical elements of correspondence analysis and multiple correspondence analysis. In: Blasius J., Greenacre M. (eds) \textit{Visualization and verbalization of data}, 31-44. CRC Press, Boca Raton

de Leeuw, J. (1973): \textit{Canonical analysis of contingency tables}, Ph.D., Leiden University, reprinted (1984) DSWO. Leiden.

Li, T.,  Ma, S. (2004). IFD: iterative feature and data clustering. In \textit{Proceedings of the 2004 SIAM International Conference on Data Mining }  472-476. Society for Industrial and Applied Mathematics.

Mackey, L. (2008). Deflation methods for sparse PCA. In \textit{Proceedings of the 21st International Conference on Neural Information Processing Systems } 1017-1024 

Mori, Y., Kuroda, M., Makino, N., (2016)   Sparse Multiple Correspondence Analysis. In: \textit{Nonlinear Principal Component Analysis and Its Applications.} SpringerBriefs in Statistics. Springer, Singapore,  47-56 

Savoy, J. (2015). Text clustering: An application with the State of the Union addresses. \textit{Journal of the Association for Information Science and Technology}, \textbf{66}(8), 1645-1654.

Shen, H., Huang, J. (2008), Sparse principal component analysis via regularized low rank matrix approximation, \textit{Journal of Multivariate Analysis}, \textbf{99}, 1015-1034.

Shen, N.M.,   Li, J.  (2015) A Literature Survey on High-Dimensional Sparse Principal Component Analysis, \textit{International Journal of Database Theory and Application}, \textbf{8}, 6 , 57-74

Shen, D.,  Shen, H., and Marron J.S. (2013) Consistency of sparse PCA in High Dimension, Low Sample Size contexts, \textit{Journal of Multivariate Analysis}, \textbf{115}, C, 317-333

Simon, N., Friedman, J., Hastie, T., and Tibshirani, R. (2013) A Sparse-Group Lasso.  \textit{Journal of Computational and Graphical Statistics}, \textbf{22}, 231-245

Trendafilov, N.T.  (2014) From simple structure to sparse components: a review. \textit{Computational Statistics}, \textbf{29}, 431–454

Trendafilov, N.T.,  Fontanella,   S.   and   Adachi,   K.   (2017). Sparse   Exploratory   Factor   Analysis. \textit{Psychometrika}, \textbf{82}(3), 778–794.

Wilms I., Croux C. (2015), Sparse canonical correlation analysis from a predictive point of view, \textit{Biometrical Journal}, \textbf{57}(5), 834-851

Witten, D., Tibshirani, R. and Hastie, T. (2009),  A penalized matrix decomposition, with applications to sparse principal components and canonical correlation analysis. \textit{Biostatistics }\textbf{10}(3),515-534.

Witten, D., Tibshirani, R. , Gross, S. , Narasimhan, B. (2019) : package PMA, \url{https://cran.r-project.org/package=PMA}

Zou, H., Hastie, T. and Tibshirani, R. (2006)  Sparse Principal Component Analysis. \textit{Journal of Computational and Graphical Statistics}, \textbf{15}, 265-286.

Zou, H., Hastie, T. and Tibshirani, R. (2007), On the “degrees of freedom” of the lasso, \textit{The Annals of Statistics}, \textbf{35}, 5, 2173–2192.

\pagebreak 
\section*{Appendix: The 45 presidents of the United States }

\begin{filecontents*}{presidents.csv}
 , , 
1. George Washington,16. Abraham Lincoln,31. Herbert Hoover
2. John Adams,17. Andrew Johnson,32. Franklin D. Roosevelt
3. Thomas Jefferson,18. Ulysses S. Grant,33. Harry S. Truman
4. James Madison,19. Rutherford B. Hayes,34. Dwight D. Eisenhower
5. James Monroe,20. James A. Garfield,35. John F. Kennedy
6. John Quincy Adams,21. Chester A. Arthur,36. Lyndon B. Johnson
7. Andrew Jackson,22. Grover Cleveland,37. Richard Nixon
8. Martin Van Buren,23. Benjamin Harrison,38. Gerald R. Ford
9. William H. Harrison,24. Grover Cleveland,39. Jimmy Carter
10. John Tyler,25. William McKinley,40. Ronald Reagan 
11. James Knox Polk,26. Theodore Roosevelt,41. George H.W. Bush
12. Zachary Taylor,27. William H. Taft,42. William J. Clinton
13. Millard Fillmore,28. Woodrow Wilson,43. George W. Bush
14. Franklin Pierce,29. Warren Harding,44. Barack Obama
15. James Buchanan,30. Calvin Coolidge,45. Donald Trump
\end{filecontents*}
\csvautobooktabular{presidents.csv}

\bigskip

\end{document}